\documentclass[preprint2,a4paper]{aastex}
\usepackage{epstopdf}
\usepackage{emulateapj5}
\usepackage{onecolfloat5}
\pdfoutput=1
\newcommand{\be}{\begin{eqnarray}}
\newcommand{\ee}{\end{eqnarray}}

\input{crab.defs}
\shortauthors{Bhat et al.}
\shorttitle{Transient detection with the GMRT}
\begin{document}
\twocolumn[
\title{Detection of fast transients with radio interferometric arrays}
\medskip
\author{N. D. R. Bhat$^{1,2,3}$, J. N. Chengalur$^4$, P. J. Cox$^{2,5}$, Y. Gupta$^4$, 
J. Prasad$^{6,4}$, J. Roy$^4$, M. Bailes$^{2,3}$, S. Burke-Spolaor$^{7,2}$, S. S. Kudale$^4$, 
W. van Straten$^{2,3}$ }
\affil{$^1$International Centre for Radio Astronomy Research, Curtin University, Bentley, WA 6102, Australia \\
$^2$Centre for Astrophysics \& Supercomputing, Swinburne University, Hawthorn, Victoria 3122, Australia \\
$^3$Australian Research Council Centre of Excellence for All-Sky Astrophysics (CAASTRO) \\
$^4$National Centre for Radio Astrophysics, Tata Institute of Fundamental Research, Pune 411007, India \\
$^5$School of Physics, University of Melbourne, Victoria 3010, Australia \\
$^6$Inter University Centre for Astronomy and Astrophysics, Post Bag 4, Ganeshkhind, Pune 411007, India \\
$^7$NASA Jet Propulsion Laboratory, M/S 138-307, Pasadena, CA 91106, USA
}
\begin{abstract}
Next-generation radio arrays, 
including the SKA and its pathfinders, will open up new avenues for exciting transient 
science at radio wavelengths. Their innovative designs, comprising a large number of 
small elements, pose several challenges in digital processing 
and optimal observing strategies. The Giant Metre-wave Radio Telescope (GMRT)
presents an excellent test-bed for developing and validating
suitable observing modes and strategies for transient experiments with future arrays. 
Here we describe the first phase of the ongoing 
development of a transient detection system for GMRT that is planned to 
eventually function in a commensal mode with other observing programs.  It capitalizes on 
the GMRT's interferometric and sub-array capabilities, and the versatility of a new 
software backend.  We outline considerations in the plan and design of transient 
exploration programs with interferometric arrays, and describe a 
pilot survey that was undertaken to aid in the development of algorithms 
and associated analysis software.  This survey was conducted at 325 and 610 MHz, and 
covered 360 deg$^2$ of the sky with short dwell times.  It provides large volumes 
of real data that can be used to test the efficacies of various algorithms and observing 
strategies applicable for transient detection.  We present examples that illustrate the 
methodologies of detecting short-duration transients, including the use of sub-arrays 
for higher resilience to spurious events of terrestrial origin, localisation of candidate 
events via imaging and the use of a phased array for improved signal detection and 
confirmation. In addition to demonstrating applications of interferometric arrays for 
fast transient exploration, our efforts mark important steps in the roadmap toward 
SKA-era science.
\end{abstract}
]

\section{Introduction} \label{s:intro}

The transient Universe has remained a major astrophysical frontier over the past few decades.
Transient phenomena are known on time scales ranging from as short as sub-nano seconds 
to years or longer,
thus spanning almost 20 orders of magnitude in time domain. Such 
emission is thought to be likely indicators of explosive or dynamic events and hence 
provide enormous potential to uncover a wide range of new astrophysics 
(e.\,g.~\citealt{cordes2004}).

While the transient sky at high energies (X- and $\gamma$-rays), and to 
some extent at optical wavelengths, are routinely monitored for transient 
and variable phenomena by a number of wide field-of-view instruments, 
it remains a largely uncharted territory at radio wavelengths.  
Most previous high-sensitivity radio surveys (for pulsars and transients) 
have used large single dishes which, by definition, have relatively narrow 
fields-of-view.  In addition, for the case of detection of short-duration
transients (``fast transients'', time scales of $\sim$microseconds to $\sim$seconds), 
there have been additional challenges such as the large signal processing 
overheads arising from the need 
to correct for effects such as dispersion, and the ever-increasing number of 
radio frequency interference sources. 
These challenges have limited the scope of rigorous explorations of the 
radio transient sky.

There are now a suite of new radio facilities in the design, construction or commissioning 
stages, many of which will offer wide field-of-view capabilities and thus open up 
new avenues of discovery. These are either multi-element radio arrays with moderate to large number 
of small-sized elements (dishes), or those comprising elements with natively wide 
field-of-view (i.\,e.~aperture arrays). Examples include the newly operational Low Frequency Array 
(LOFAR) and the  Murchison Widefield Array (MWA), as well as upcoming SKA pathfinder instruments, 
{\it viz.} the Australian SKA Pathfinder (ASKAP) in Western Australia and MeerKAT in South Africa \citep{lofar2011,mwa2012,askap2007,meerkat2009}. 
In principle, these instruments can provide large field of view (FoV) observations; 
however, they also present significant 
challenges in terms of the associated signal processing costs. Fortuitously, with the 
recent advances in affordable super computing and the use of graphics processing units 
in astronomical computing, this is fast becoming less of a challenge 
\citep[e.g.][]{barsdelletal2010,magroetal2011}. Therefore, optimistically, the availability 
of such next-generation arrays, together with appropriate instrumentation and suitable 
data archiving and processing strategies, can potentially revolutionize our knowledge of 
the transient radio sky in the coming decades. 

The scientific potential of radio transients has been well underscored in a number of 
recent reviews \citep[e.g.][]{cordes2004,cordes2009,fender-bell2011,bhat2011}. 
A wide variety of transient phenomena are known at radio wavelengths. While 
pulsar radio emission time scales range from milliseconds 
(sub-pulses) to nanoseconds (giant pulses), phenomena such as solar or 
stellar bursts, flares from Jupiter-like planets and brown dwarfs, micro-quasar emission, and gamma-ray burst (GRB) afterglows are of 
much longer durations \citep[e.g.][]{chandra-frail2011}. 
Some known radio transients 
have been discovered in follow-up observations of higher-energy detections; for example, 
gamma-ray burst afterglows and periodic pulsations from magnetars 
\citep{camiloetal2006,levinetal2010}. Other discoveries include transient 
sources in the direction of the Galactic Centre (GC) \citep{hymanetal2005,boweretal2007,roy2010} 
found through time-resolved VLA imaging of the GC, rotating radio transients (RRATs) 
found in transient searches of archival pulsar surveys \citep{rrats2006,km2011}, 
and the possibly extragalactic millisecond bursts reported by \citet{lorimeretal2007} and \citet{keaneetal2012}. 

A distinction is often made between ``slow'' versus ``fast'' transients in the context of radio astronomy 
\citep[cf.][]{cordes2009}; 
{\it slow} transients can be detected through standard imaging of brief or long time integrations, 
while {\it fast} transients require data collection with sufficiently high time and frequency resolution to correct for dispersive delays before detection is attempted.
This paper is concerned with the detection of fast transients. These are often linked to coherent radiation processes and, frequently, to sources 
in extreme matter states
\citep[e.g.][]{cordesetal2004}. They are affected by plasma propagation 
effects such as dispersion and, if the source is compact, by multi-path scattering 
and/or scintillation by the intervening media; hence, they may also serve as 
excellent probes of such media.

As noted earlier, impulsive radio frequency interference 
(RFI) can potentially mimic signatures of real signals, and their frequent occurrence may 
impact an observation's sensitivity, thereby making weaker signals difficult to detect 
\citep[e.g.][]{bhatetal2005}. Interferometric instruments offer several unique 
advantages here.
The distributed nature of array elements and long baselines can be
exploited to identify and eliminate a wide range of RFI-generated transients.
For example, voltage data can be correlated between elements to find fringes for the pulse, hence obtaining a sky position and localizing the detection.
Most ongoing fast transient explorations, with the exception of the VLBA-based V-FASTR project \citep{waythetal2011} and the LOFAR pulsar survey project \citep{lofar2012}, use large single-dish instruments such as Parkes and Arecibo \citep{denevaetal2009,bsetal2011}, which offer none of those advantages, both because of their lower resilience to RFI and also because the data are typically pre-processed prior to recording. 

Despite the clear advantages of interferometric transient searches, exploiting such arrays will 
require considerable planning and exploratory research.  As neither of the conventionally employed 
observing strategies, such as incoherent (i.e. phase-insensitive) addition of antenna signals or 
a single phased-up array, are optimal for conducting large sky surveys, some new strategies 
will need to developed and experimented in order to fully exploit 
array instruments  \citep[e.g.][]{gemma2009,lofar2011,lofar2012,rubio2013}. Recently, \citet{jp11} and \citet{cc11} 
approached the problem from the 
point of optimizing large-sky surveys within the context of next-generation array instruments 
including the SKA, and both advocate incoherent combination of antenna signals as optimal 
strategies to achieve the highest detectable event rates. Existing arrays (e.\,g.~GMRT, VLBA, LOFAR) 
can meanwhile demonstrate effective strategies that will be applicable when next-generation 
arrays are constructed.

A number of salient features make the GMRT \citep{swarupetal1991} a powerful test-bed in 
this context. 
This low-frequency array of 30 x 45-m dishes, operating at 5 different frequency bands in the 
range 0.15 to 1.5 GHz and with an effective collecting area ${\rm A_{eff}}\sim$3\% SKA, 
offers several unique design features. 
Its moderate number of elements, relatively long baselines (up to $\sim$25 km) 
and sub-array capabilities make it an excellent analog for 
SKA-like platforms. Furthermore, GMRT's new software backend \citep{royetal2010}
allows raw voltage data from individual array elements to be rerouted to software-based processing systems.

\begin{figure*}[t]
\epsscale{2.0}
\plottwo{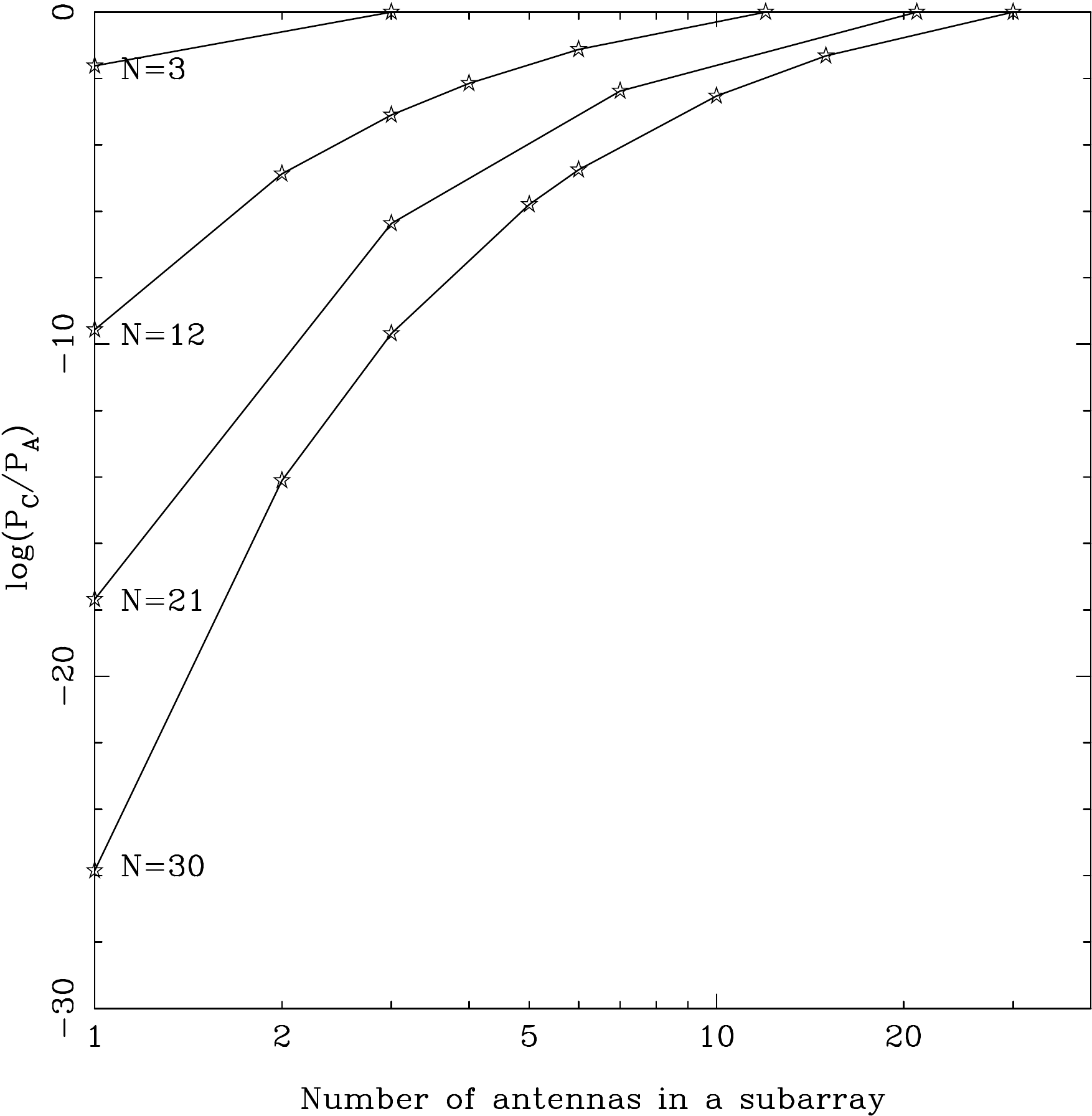}{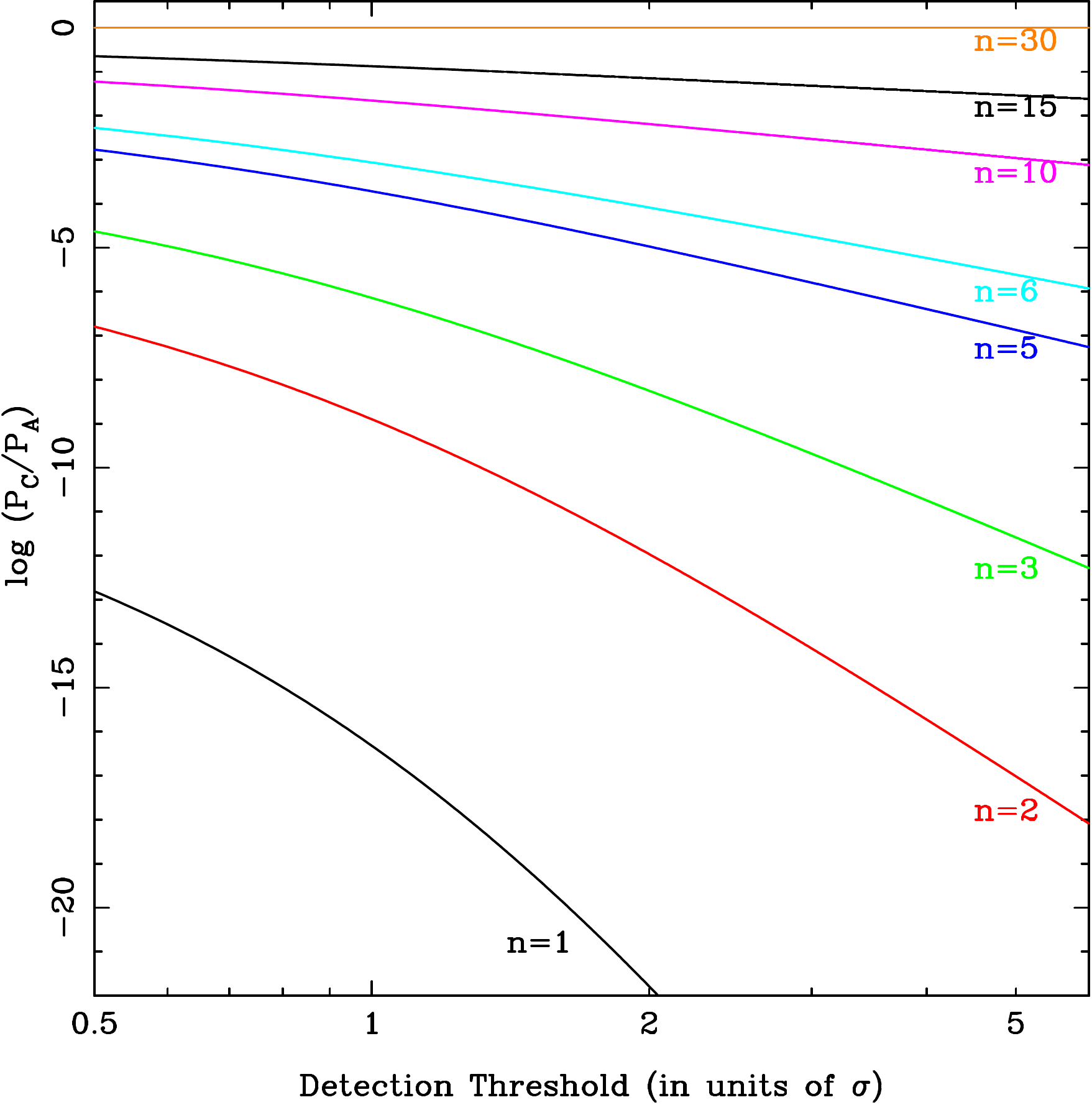}
\caption{Ratios of the probabilities of false alarms, $P_{C} / P_{A}$, where $P_{A}$ 
is the probability for a sub-array where the signals from all the $N$ antennas are added
incoherently, and $ P_{C} $ is the joint probability for the case of $p$ sub-arrays 
each having $n = N/p$ antennas. 
{\it Left panel}: The ratios are plotted as a function of the number of antennas in a 
sub-array and for a detection threshold $ T / \sigma $ = 3, for the total number of 
antennas $N$ = 3, 12, 21 and 30 respectively. 
{\it Right panel}: the same quantity $P_{C}/P_{A}$ is now plotted as a function of the 
threshold (in units of $\sigma$) for various sub-array combinations ($n$ = 30, 15, 10,
6, 5, 3, 2, 1) from the top to the bottom curve, for the case of $N ~=~ 30$ (note that 
the curve for $n$ = 30 corresponds to the top horizontal line at log $P_{C}/P_{A}$ = 0) 
-- see the text for more details. } 
\label{fig:jayanti}
\end{figure*}

Here we will describe ongoing efforts to equip the GMRT for transient exploration 
by (i) designing a software based system that will eventually function commensally
with other observing programs, and (ii) undertaking pilot surveys that help demonstrate 
observational methodologies. This paper will focus on algorithms and methodologies, 
while the  detailed implementation of a real-time processing 
pipeline and science results from pilot surveys are deferred to future papers. Apart 
from demonstrating the application of a ``large-N, small-D" (LNSD) type instrument for 
transient explorations, these efforts will also enable new science with the GMRT. 
This is especially important given that the GMRT transients surveys will complement 
other similar efforts around the world in sky and frequency coverage.

This paper is organised as follows. In \S 2, we outline considerations that drive 
transient exploration strategies with interferometric instruments. In \S 3 we 
highlight unique advantages of the GMRT for this topic,
and describe pilot surveys undertaken to aid the necessary technical development. 
Details of our transient detection pipeline are discussed in \S 4, and applications to real 
data are presented in \S 5. In \S 6 we discuss our event analysis pipeline and present 
examples illustrating important methodologies. In \S 7 we comment on possible future 
directions and in \S 8 we present our conclusions.

\section{Interferometric Arrays for Transient Searches: Considerations and Strategies} \label{s:cons}

In this section we discuss various considerations in searching for fast transient signals with 
interferometric instruments. We discuss various technical and sensitivity 
considerations that arise from the distributed nature of array elements, the role of propagation 
effects in signal detection and analysis,  the importance of searching over a large 
parameter space and the use of long baselines to serve as spatial filters against RFI.
While much of our discussion is presented within the context of the GMRT, we emphasise 
that these discussions are also applicable to other similar, particularly low-frequency, array 
instruments.

\subsection{Technical and Sensitivity considerations} \label{s:tech}

Array instruments can be used either in ``incoherent array'' (IA) or ``phased array'' (PA) modes for 
time-domain applications such as
observing pulsars \citep{guptaetal2000}, and in principle, similar strategies can be considered
for the detection of fast transients. IA and PA correspond to modes which maximise the FoV 
and detection sensitivity (or the effective 
collecting area ${A_{\rm eff}}$), respectively.
The IA mode is good for surveys, however it comes at the expense of a significant reduction 
in overall sensitivity. At the other 
extreme is a fully coherent array mode, where the signals from individual elements have to 
be combined to produce (many) phased-array beams within 
the primary beam in order to achieve the full FoV of the single element. This can be 
prohibitively expensive in terms of the real-time signal processing costs, as the number of 
beams goes as $(D/d)^2$, where $D$ is the physical extent of 
the array and $d$ is the size of the individual element or dish.  For instance, application 
to just the central square (1\,km $\times$ 1\,km) of the GMRT requires the formation of 
$\sim$500 beams, whereas over $\sim10^5$ beams will be required in order to realise the 
full FoV and sensitivity of the array. As a general rule, the use of phased-array beams for 
large surveys becomes less appealing as the filling factor of the array starts to fall-off.

An intermediate strategy that tries to optimize the trade-off between sensitivity and
FoV and to maximise ${\rm A_{eff} \times FoV}$, while offering
additional advantages for transient searches, is to use distinct sub-arrays with appropriately 
combined signals.
These sub-arrays could be incoherent or coherent formations, 
for which we may then use statistical measures on sub-array detections to optimise the 
performance with respect to sensitivity, FoV, radio frequency interference (RFI) and 
excision of false
positives etc. Here we describe the basis for such a scheme that has been implemented
and tested using the GMRT array.  

Our basic strategy is to generate a small number of incoherently summed
sub-arrays and combine the candidate transient event detections from the
sub-arrays in a manner that optimises the rejection of false positives via 
suitable coincidence filtering techniques. This will preserve the full FoV 
of a single element.  
To motivate this strategy, we consider the probability of false alarms
in a transient detection scheme for various combinations of sub-arrays made from
an array of N antennas.  As described in detail in the Appendix, for an array of 
$N$ elements configured to make $p$ subarrays (with \mbox{$n=N/p$} elements per 
sub-array), the joint false alarm probability is given by

\begin{eqnarray}
P_{C} ( > T )  = \left[ \frac{1}{2} \mbox{Erfc} \left( \sqrt{\frac{N}{2p}} \frac{T}{\sigma} \right) \right]^p~,
\label{eq:prob}
\end{eqnarray}
where $T = r\,\sigma$ is the detection threshold and Erfc is the complementary error function (i.e. $r$ is the detection threshold in units of $\sigma$).
As shown in the Appendix, $P_C$ can be significantly less than $P_A$, the false alarm probability for a single sub-array \mbox{($p=1, n=N$)}.
This is illustrated in 
Fig.~\ref{fig:jayanti}, which shows the ratio $P_C/P_A$ as a function of $n$ for 
different cases of $N$ and a fixed value of \mbox{$r=3.0$} (left panel), and as a function of
$r$, for different choices of $n$ and a fixed value of \mbox{$N=30$} (right panel).  

As evident from these figures, for an array of 30 antennas like the GMRT, the false detection rate can be improved by a few orders of magnitude by splitting the array into 4-5 sub-arrays
of 6 to 7 antennas each.  It is also clear that the improvements increase with
a greater value of detection threshold.
Since $r$ (as defined in the Appendix), is relative to $\sigma$ for the signal from a single antenna, realistic values (e.\,g.~a 5-$\sigma$ threshold) for the different array combinations correspond to values
of \mbox{$r<5$} (e.\,g.~\mbox{$r\sim 1$} corresponds to a $\sim$5-$\sigma$ detection threshold for a single 30-antenna sub-array).  For this range of $r$ values, 
improvements in the false detection rate by a factor of 10--100 can be obtained 
by splitting the array into 4-5 subarrays, while using the same detection threshold
(say 5-$\sigma$). 

Note that for the sub-array case, operating at the same detection threshold corresponds 
to a lower absolute sensitivity than the full array case.  However, it should be possible to trade-off the false positive rate (while still keeping 
it below or comparable to that for the full array case) by reducing the threshold 
appropriately, thereby increasing the absolute sensitivity 
and bringing it closer to that of the full array. 
The sub-array case is expected to offer other advantages that accrue from rejection of false positives,
such as discriminating against localised RFI; the effectiveness of this would 
depend on the physical extent of the full array, and how the antennas are grouped to 
form the sub-arrays.

\begin{figure}[b]
\epsscale{0.99}
\plotone{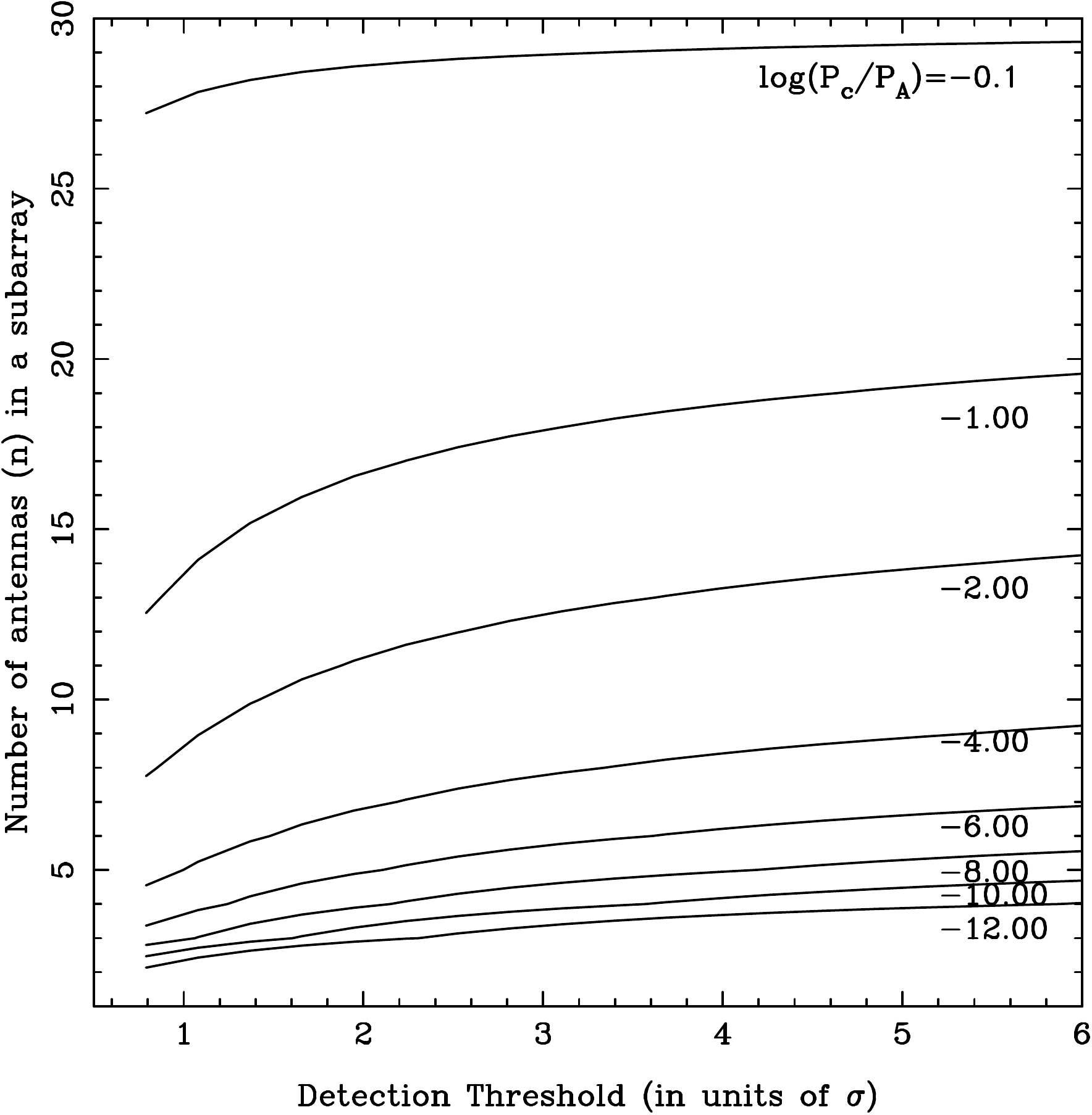}
\caption{The false alarm probability when sub-arrays with coincidence filters are 
used for the rejection of false positives. The resultant probability, $P_C$, is normalized 
to the probability for a single, 30-antenna incoherent array, $P_A$, and is shown as 
a function of the detection threshold $r$ and the number of antennas per sub-array $n$. 
Curves are drawn for ${\rm log} (P_C/P_A) = -0.1, -1, -2, -4, -6, -8, -10, -12$ (top to bottom). 
These can be used as a guide to make the choice in terms of number of sub-arrays for 
a set threshold and desired level of rejection of false positives. For example,  a 2$\sigma$
threshold ($r$=2) and $P_C/P_A$=100 require dividing the array into 10 sub-arrays.}
\label{fig:cont}
\end{figure}

Fig.~\ref{fig:cont} shows 
$P_C$ after normalisation to the probabilities of false alarms
for a single incoherent array of all 30 antennas. Such plots may serve as 
useful guides to design an optimal observing strategy, e.\,g.~to determine the number of sub-arrays required to realize a desired false positive rate for a set threshold, or to determine the threshold value that will be needed to achieve the desired level of rejection for a chosen number of sub-arrays.

The absolute sensitivity considerations are as follows.  The sensitivity of a single 
element is characterised by its gain ($G_a$) and the system temperature (\Tsys). 
For a sub-array of N antennas, a signal is detectable 
if its peak flux density (\Speak) exceeds some minimum flux density as determined by the 
radiometer equation:

\begin{equation}
S_{\rm pk,min} \, = \, K \, {\beta \, (T_{\rm rec} + T_{\rm sky}) \over G_{\rm n} \, (\Delta \nu \, N_{\rm pol} \, W_{\rm p}) ^{1/2} }~,
\label{eq:sens}
\end{equation}

\noindent
where $T_{\rm rec}$ and $T_{\rm sky}$ are the receiver and system temperatures, 
respectively ($ T_{\rm sys} \approx T_{\rm rec} + T_{\rm sky} $ for most instruments),
$G_{\rm n}$ is the net gain of the array in ${\rm K \, Jy^{-1}}$,  $\Delta \nu$ is the recording 
bandwidth, $N_{\rm pol}$ is the number of polarisations, $W_{\rm p}$ is the 
matched filter width employed in transient searching, $\beta$ denotes the loss in 
signal-to-noise ratio (S/N) due to signal digitization, and the factor K is the detection threshold 
in units of rms flux density ($\sigma$). 


For a 30-antenna sub-array on the GMRT, operating with a bandwidth of
32 MHz for a 5-$\sigma$ threshold, the achievable sensitivity for a survey would 
typically range from $\sim$1 Jy (at 610 MHz) for $W_p$=1 ms to $\sim$0.1 Jy for $W_p$=100 ms. 
As the single-antenna gain ($G_a$) and \Tsys (when pointed to the cold sky) are comparable at 327 and 610 MHz, the nominal sensitivities are similar at both frequencies, 
though in practice the larger 327\,MHz sky background (\Tsky) will degrade 
the sensitivity. 

\begin{figure}[b]
\epsscale{1.0}
\plotone{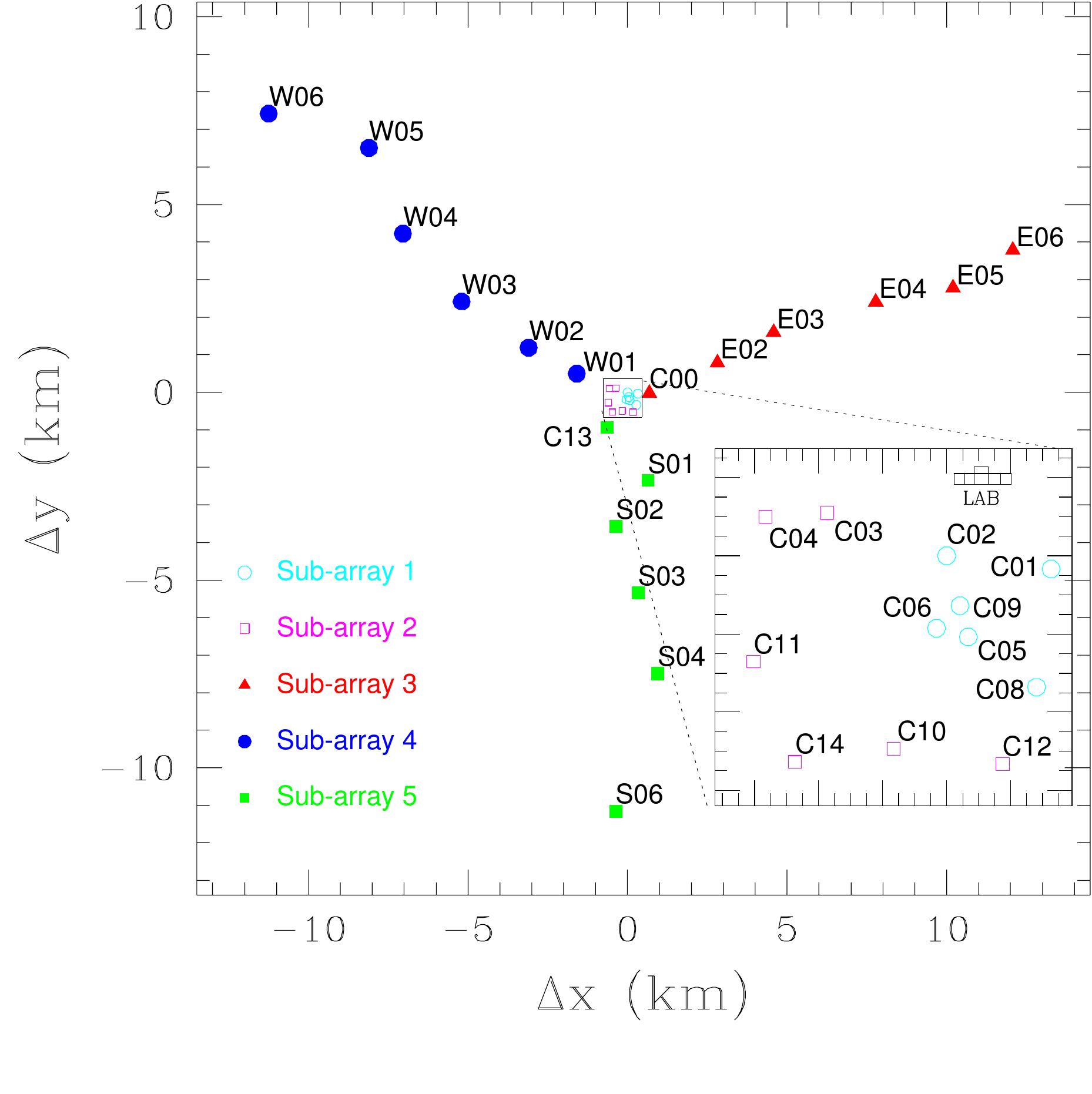}
\caption{
A map of the GMRT array indicating antenna locations across the east, west
and south arms and in the central 1\,km\,$\times$\,1\,km region. The offsets 
$ {\rm \Delta x } $ and $ {\rm \Delta y } $ are relative to the antenna C02, 
which is at ($ {\rm \Delta x } $, $ {\rm \Delta y } $) = (0, 0). 
The antennas can be grouped into multiple distinct sub-arrays of nearly equal 
sensitivities. The case for five sub-arrays is shown here: with three sub-arrays
formed from antennas across the three arms, and the other two from antennas 
located in the central square. Such arrangements can yield high efficiencies in 
the identification and elimination of spurious events due to RFI.} 
\label{fig:layout}
\end{figure}

When using sub-arrays (see e.g., Fig.~\ref{fig:layout} where the array is divided into five 
sub-arrays of six antennas each),  however, $G_{\rm n}$ scales as $\sqrt{n} \, G_{\rm a}$.
This leads to a worse sensitivity than that for a single
$N$-element sub-array by a factor of $\sqrt{p}$.  As mentioned above, some or all of
this loss can be recovered by lowering the threshold by a corresponding 
factor, provided the resultant false positive rate remains better than that achievable 
for the default threshold with the single sub-array case.   
 
\subsection{Propagation effects} \label{s:prop}

The role of plasma propagation effects in fast transient detection is discussed 
in detail by \citet{cordes2009} and \citet{jp11}. These include dispersion, 
pulse broadening or scattering, and scintillation, and they are due to ionised 
interplanetary, interstellar and/or intergalactic media. While for most Galactic 
sources, the dominant contribution is from the interstellar medium (ISM), for 
sources at extragalactic or cosmological distances, there may also be significant 
contributions from the ISM of the host galaxy as well as from the intergalactic 
medium (IGM). 

The differential dispersion delay, $\Delta t  _{\rm dm} $ (in ms), across an observing 
bandwidth $\Delta \nu$ centred at an observing frequency $\nu$ (both in GHz) is given by 
$ \Delta t  _{\rm dm} \approx 8.3 ~ {\rm DM} \, \Delta \nu \, \nu^{-3} $, where DM is the 
dispersion measure. For Galactic sources, DM can be up to several thousand \dmu at large 
Galactic distances or toward the GC. Away from the Galactic plane, such large 
DMs can be expected for signals of extragalactic or cosmological origins.

\begin{figure}[t]
\epsscale{1.0}
\plotone{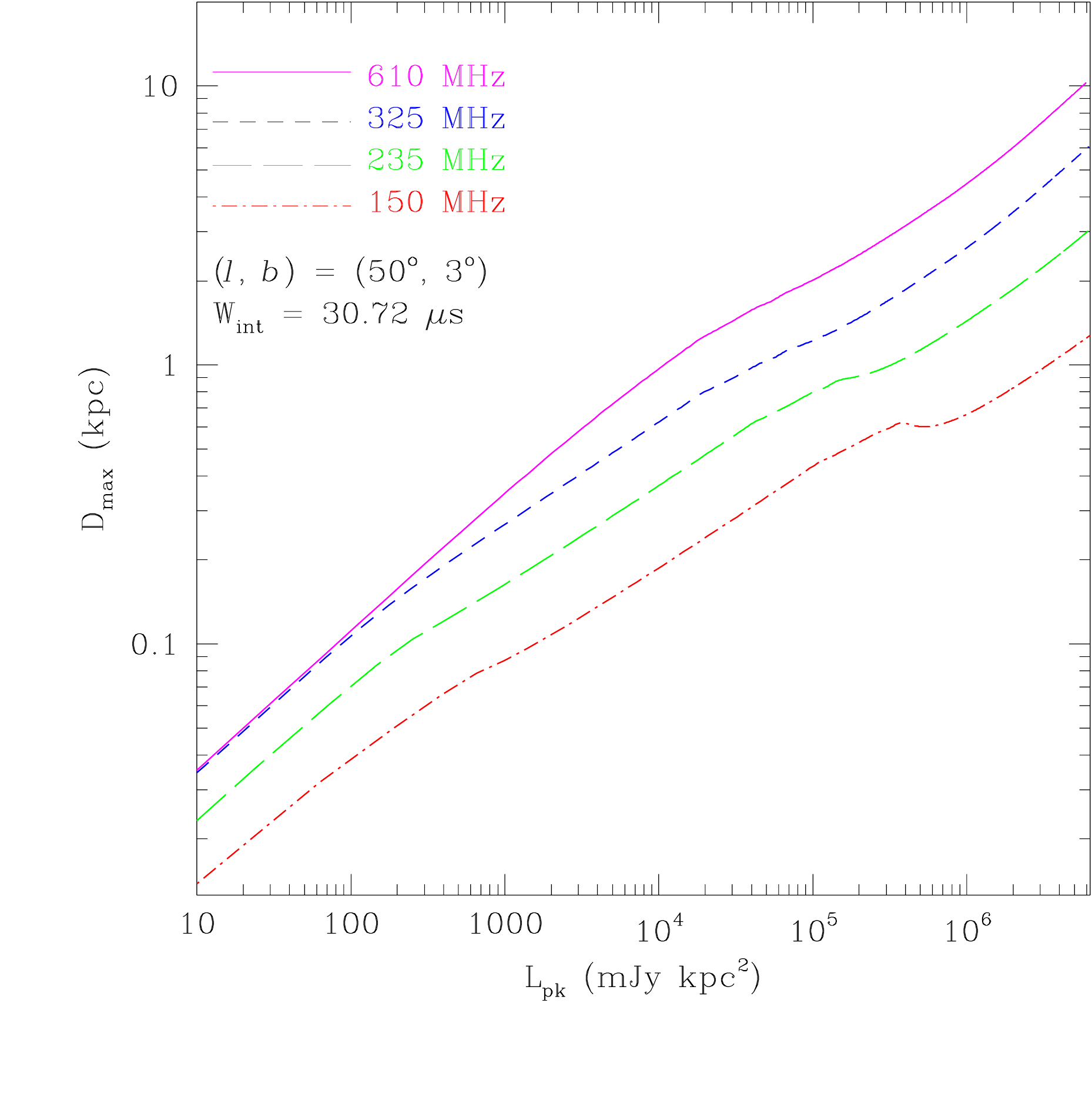}
\caption{Plots of maximum distance to which transient detections are possible ($D_{\rm max}$
vs peak luminosity $L_{\rm pk}$), for the GMRT's low frequency bands. $D_{\rm max}$ tends to 
vary linearly at low luminosities (i.\,e.~smaller distances) when scatter broadening is negligible,
and more slowly with increasing luminosity at large distances when scattering becomes 
prominent. While the nominal sensitivities are comparable for the GMRT at 325 and 610 MHz, 
scattering is important even at relatively lower distances at 325 MHz. The effect is 
more pronounced at lower frequencies, resulting in significantly lower values
of $D_{\rm max}$ for $L_{\rm pk}$ and hence smaller search volumes for detectable 
signals.} 
\label{fig:dmax}
\end{figure}

Scattering (pulse broadening) leads to asymmetric pulse shapes with a stretched pulse tail. 
Measured pulse broadening times  (\taud) scale steeply with  the observing frequency;
$ \taud \propto \nu^{-3.9\pm0.2}$ from observations \citep{bhatetal2004}. 
Detection will thus become difficult when $\taud\gg\Wint$, the intrinsic width of emission.
For $\taud \ga \Wint$, signal detection can still be critically influenced by the degree 
of scattering. While pulse broadening conserves the fluence (i.\,e.~integrated flux), the 
smearing in time leads to smaller pulse amplitudes (i.\,e.~lower peak flux densities), 
and hence lower signal-to-noise in the detection. 
Scattering can thus play an important role in 
defining optimal search strategies with low frequency arrays such as the MWA and 
LOFAR as well as the GMRT.

Both diffractive and refractive effects are  important at low frequencies. 
Diffractive scintillation  produces structure in both time and frequency, 
with the characteristic scales $\sim$100\,s in time and $\sim$100\,kHz in frequency for 
observations made at $\sim$300--600\,MHz and for DMs \la 50 \dmu \citep[e.g.][]{guptaetal1994,bhatetal1998}. 
As diffractive time scales 
are typically longer than $\sim$seconds, an apparent brightening or dimming of 
signals may arise in cases where diffractive bandwidth ($\nu_d$) is of the order 
of, or larger than, the recording bandwidth ($ \nu _d \, \ga \Delta \nu $).  For 
distant sources, $\nu _d  \ll \Delta \nu $, thus signal 
detection will be minimally affected. Refractive scintillation, on the other hand, 
leads to slow flux modulation on time scales of $\sim$days to weeks 
or longer \citep[e.g.][]{guptaetal1993,bhatetal1999}.

Regardless of their impact on signal detectability, propagation effects 
can potentially serve as a useful discriminator from local RFI.
While it may be possible for certain types of RFI to mimic one or 
more propagation effects, it is unlikely that non-astrophysical signals will emulate 
multiple effects in a manner consistent with models of astrophysical media. 

\subsection{Parameter space and search volume} \label{s:para} 

In searching for short-duration radio transients, the two most basic search parameters are: 
(i) DM, and (ii) the duration of the signal. The latter is typically quantified as 
the effective pulse width, \Wp. Here we briefly discuss the 
search parameter space, particularly in terms of limitations imposed by dispersion, scattering, detection sensitivity, and search volume. 


\begin{figure*}
\epsscale{2.0}
\plotone{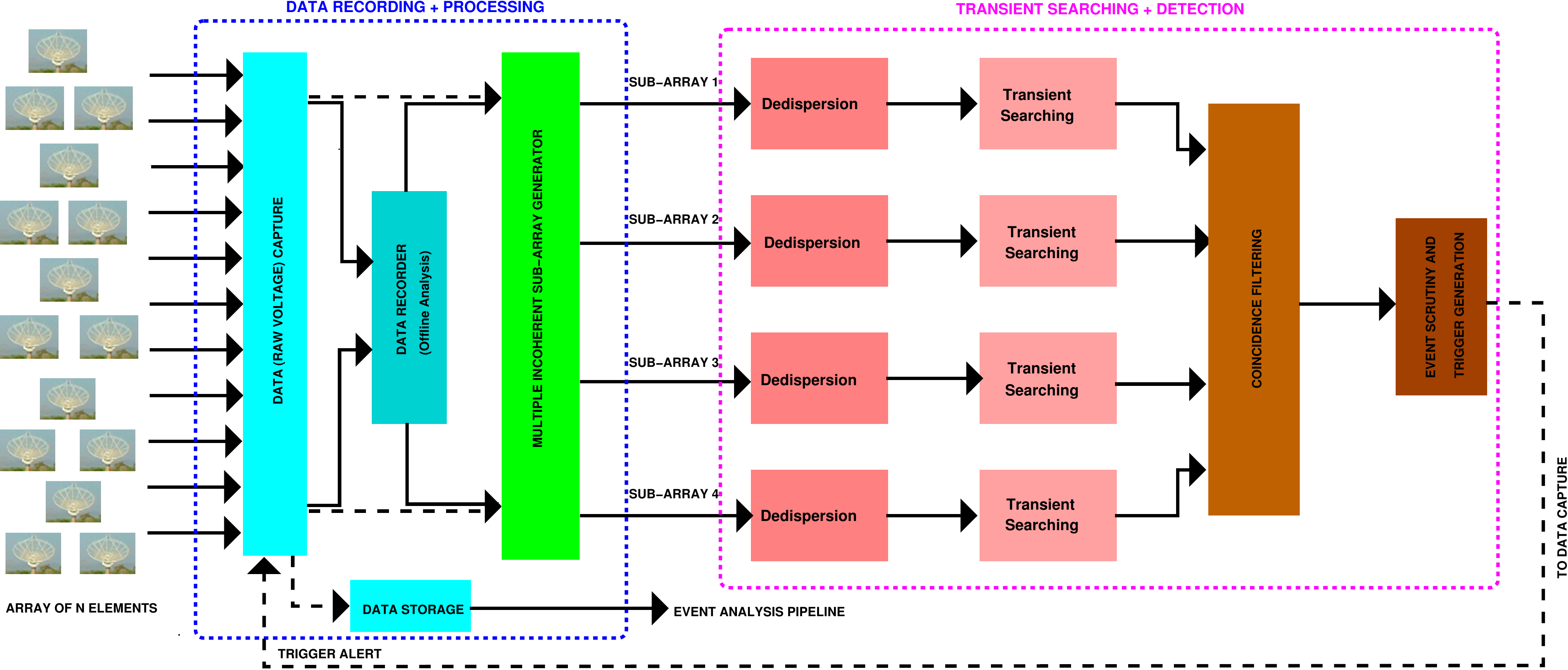}
\caption{
A diagrammatic representation of the GMRT transient detection pipeline. 
Raw voltage data from each array element are captured and made available 
to a processing pipeline. Multiple (incoherent) sub-array data streams are
generated, and a transient search is performed on each data stream. 
The resulting events are assimilated through the event identification and 
coincidence filter algorithms to select the final candidates. For real-time 
implementation (dashed lines and arrows), this information is then used 
to generate triggers that will alert the raw data capture system to record 
relevant raw data segments for further detailed processing and scrutiny.
} 
\label{fig:pipeline}
\end{figure*}

As discussed above, dispersion delays can be substantial, even at moderate 
DMs, for low radio frequencies; e.\,g.~a  pulse with $\Wint=1$\,ms 
and DM = 10 \dmu will be smeared over $\sim$100\,ms in observations 
with $\Delta\nu=32$\,MHz centered at 300\,MHz. While it is generally 
advisable to search out to very large DMs, in practice for searches within 
or near the Galactic plane, the maximum DM that can be effectively 
searched will likely be limited by pulse broadening.  As the number of 
trial DMs are typically determined from analytical constraints that ensure 
minimal degradation of S/N due to DM errors, the DM spacings tend to be 
fairly small at low frequencies, thereby requiring a large number of trial 
DMs to span a given DM range. This can translate to significant processing 
costs for low frequency searches.


\begin{figure*}[t]
\epsscale{1.5}
\plotone{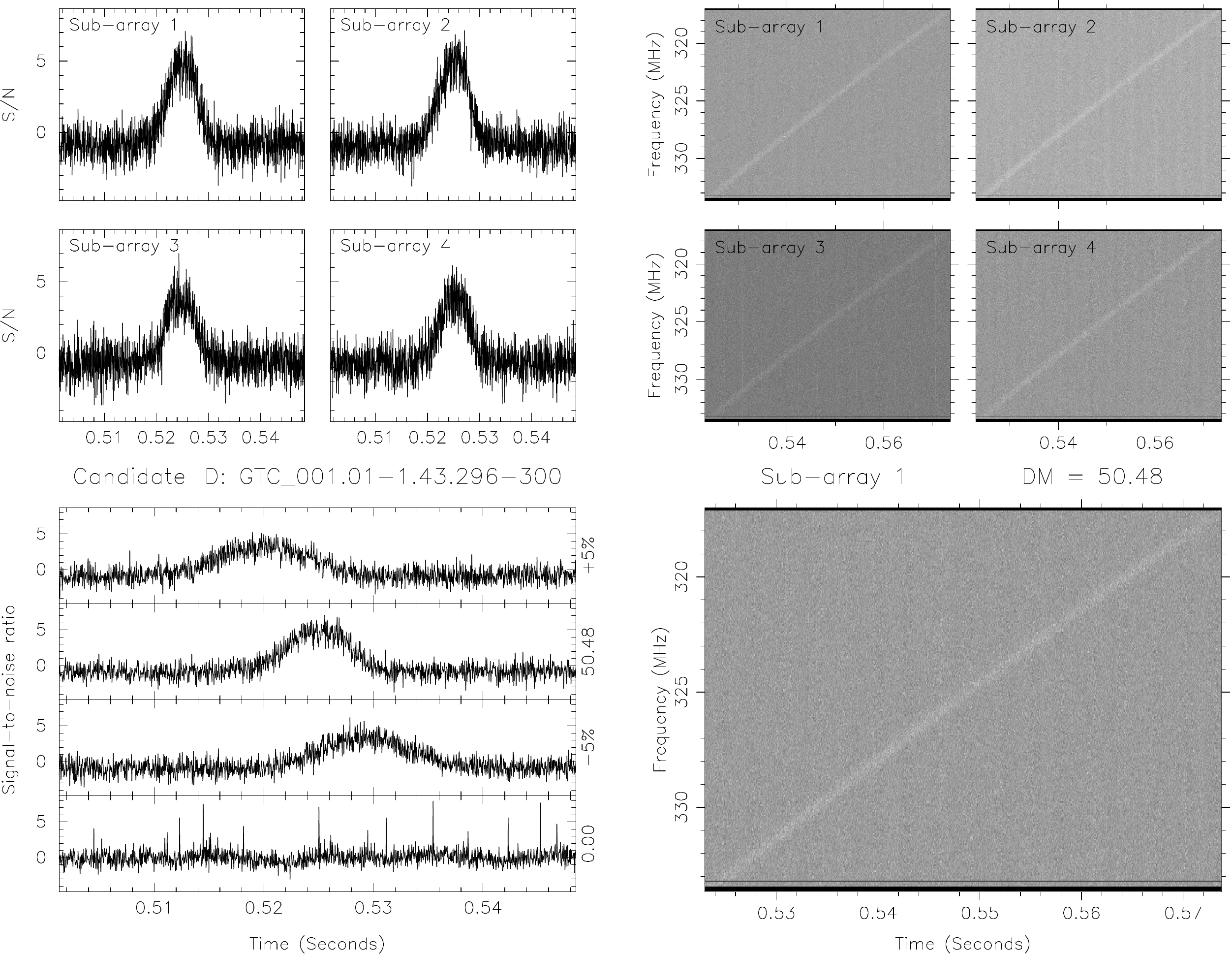}
\caption{Example plots from the GMRT transient detection pipeline.  A transient pulse 
was detected in the survey field GTC\_001.01--1.43 (observations at 325 MHz, i.\,e.~beam 
width $\approx$80$^{\prime}$). The pulsar PSR J1752$-$2806 is located at an offset of 
42$^{\prime}$ from the phase center. The array was divided into four groups of roughly 
equal antennas, however the detection significance for the sub-arrays 2 and 3 (i.\,e.~the 
south arm and the central square) were relatively lower, possibly due to the failure of some 
antennas to function at their nominally expected sensitivities. The top sub-panels show the 
de-dispersed time series (left) and frequency-time excerpts around the detected event 
(right). The bottom left sub-panels show the signal at two nearby DMs as well as at zero 
DM in addition to the candidate DM, while the bottom right panel is an enlarged version 
of the frequency-time plane excerpt around the signal. The detection of the signal in all 
four sub-arrays, its broad-band nature, dispersion sweep and a reduced S/N at nearby 
DMs and the absence of signal at DM=0 serve as multiple positive detection diagnostics.}
\label{fig:example}
\end{figure*}

\begin{figure*}[t]
\epsscale{1.5}
\plotone{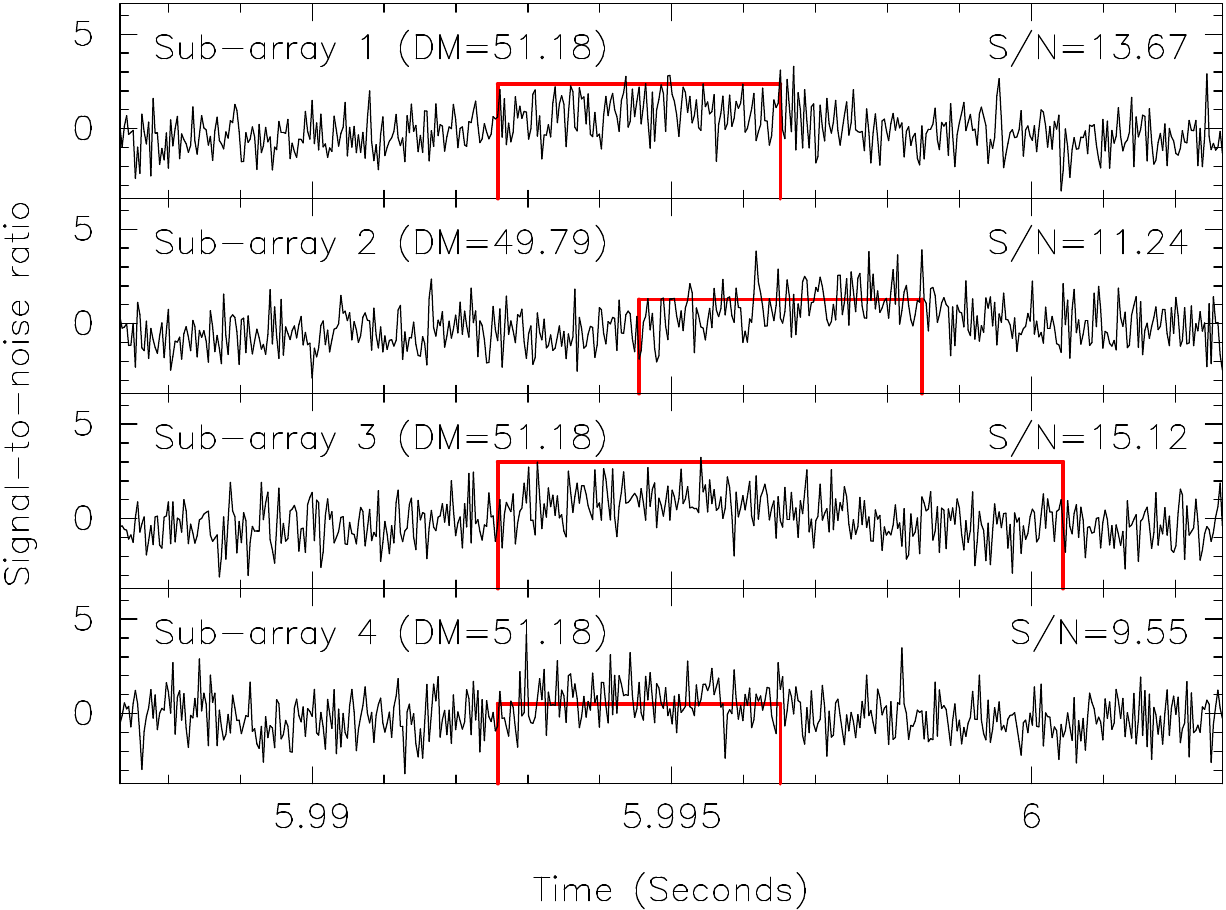}
\caption{Another example detection from the GMRT transient pipeline, where the noise fluctuations
lead to a transient pulse being detected at slightly different DMs and pulse widths in addition 
to different peak S/Ns across different sub-arrays. The true DM 
of the signal (i.\,e. a pulse from PSR~J1752$-$2806 at $\sim71^{\prime}$ offset from the phase 
centre) is closer to the DM value reported by the sub-array 2, in which the pulse was detected at a 
slightly reduced S/N (by $\sim10\%$) at DM=51.18 \dmu (i.\,e. the best DM as reported by 
other three sub-arrays).  The widths and heights of the rectangle (red) boxes are proportional 
to the effective widths and peak S/Ns as found by the processing pipelines.}
\label{fig:weakpulse}
\end{figure*}

The vast spread in the duration of known transient phenomena 
make a compelling case to search in time duration over as wide a range
as possible. In practice, the shortest time scale 
that can be effectively searched is limited to the sampling interval 
achievable with the recording instrument ($dt$); any signals of $\Wint \la dt$ will thus be
instrumentally broadened to $dt$. However, at low frequency, pulse broadening of astrophysical 
origin will likely exceed instrumental broadening, even at moderate DMs.\footnote{For instance, $\Wint\la1\,\mu$s emission from the Crab is broadened 
to $\sim$100$\,\mu$s at 600\,MHz and $\sim$1\,ms at 300\,MHz \citep{bhatetal2007}.}  At 
large DMs however, pulse broadening will limit the achievable (effective) time 
resolution, and it will be difficult to detect heavily scattered pulses owing to 
S/N degradation from broadening. The longest time durations that can be searched 
will therefore be dictated by pulse broadening. 


\begin{figure*}[t]
\epsscale{2.0}
\plotone{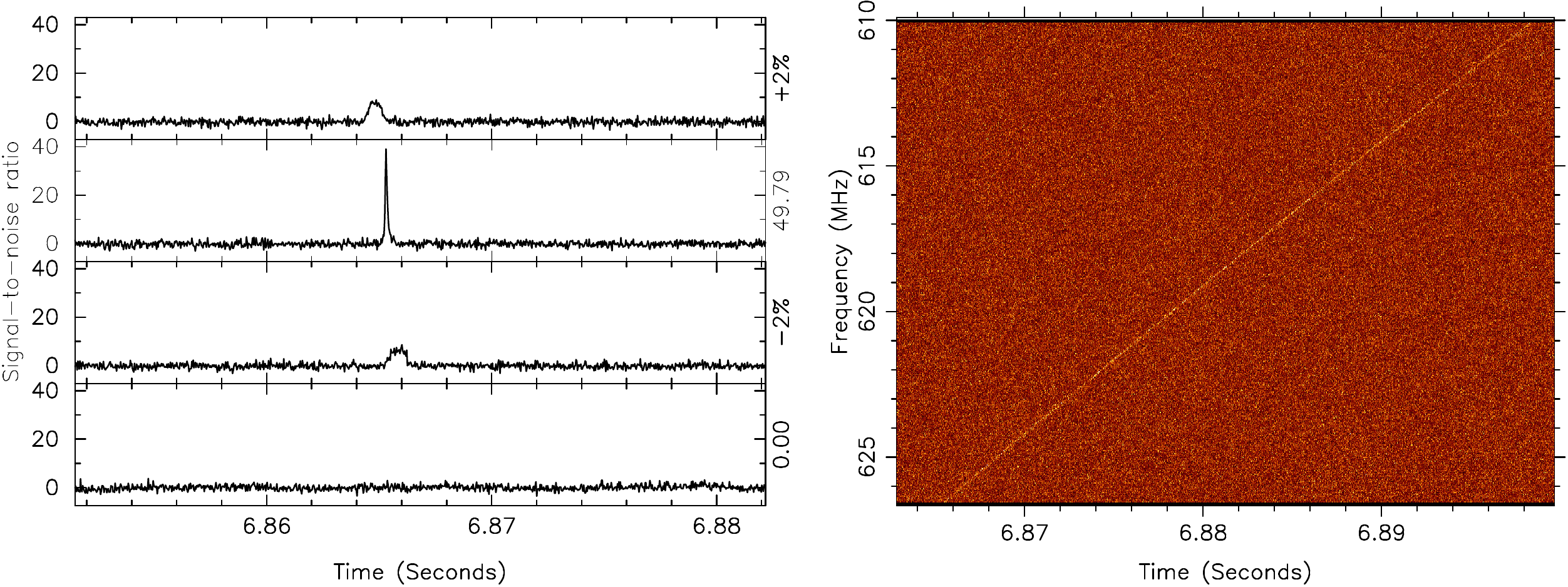}
\caption{Example plots from the GMRT transient detection pipeline -- 
the array was configured into seven different sub-arrays, each comprising a single antenna, 
thus providing a powerful coincidence filter against spurious events of RFI origin. 
Data were collected by emulating a `survey mode', by scanning the 
sky region around the Crab pulsar at 0.5 ${\rm deg~min^{-1}}$. A bright giant pulse 
was detected as a `transient' when the pulsar was within the telescope beam 
(half power beam width $\sim$0.5$^{\circ}$). The pulse is very narrow 
($\approx$50  $\mu$s), approximately twice the sampling resolution (30 $\mu$s) 
and so is seen as a thin strip on the waterfall plot. The signal  peaks at 
DM = 56.74\,\dmu, with a sharp decline in S/N even at small departures from the 
true DM; for instance, ${\rm \Delta DM / DM \approx 0.02}$ results in a S/N loss 
of almost a factor of four, exemplifying the need for very short DM spacings and 
high time and frequency resolutions for transient searches at low radio frequencies.}
\label{fig:crabgiant}
\end{figure*}

Propagation effects may also significantly influence the maximum distance to which 
a detection is possible, $D_{\rm max}$, and therefore the search volume, given by 
$V_{\max} = (1/3) \Omega_{\rm s} D_{\rm max}^3$, where $\Omega_{\rm s}$ is the 
FoV. As $D_{\rm max}$ scales as $S_{\rm pk,min}^{-1/2}$, 
the prominent low-frequency effect is pulse broadening. The resultant 
amplitude degradation ($S_{\rm pk} \propto \tau _d ^{-1/2}$) leads to a lower 
$D_{\rm max}$ and consequently a smaller search volume. A detailed treatment of 
this effect and relevant survey metrics are given by \citet{cordes2009}, who considers 
different possible survey strategies, both for fast and slow transient searches 
with the SKA. Following the formalism presented there, useful plots can be made 
of $D_{\rm max}$ vs $L_{pk}$ (where $L_{pk} = S_{\rm pk} \, D^2 $ is the peak 
luminosity) for a given choice of search parameters.
As an illustration, Fig.~\ref{fig:dmax} shows such sensitivity plots for different 
GMRT frequencies, for one specific line of sight within our pilot survey 
region ($l$=50$^{\circ}$, $b$=3$^{\circ}$). Here we account for various 
propagation effects, instrumental broadening, and the increase in 
sensitivity from using matched filtering. Reduced sensitivities (in the lower 
$L_{pk}$ range) at 150 and 235 MHz are due to the relatively 
larger sky backgrounds at these frequencies. As evident from these plots, 
$D_{\rm max}$ is reduced at higher  $L_{pk}$  (i.\,e.~larger distances 
for a given $S_{\rm pk}$), resulting in departures from linear trends compared to 
the lower $L_{pk}$ range. This effect is obviously direction dependent, thus 
making detection rates a strong function of sky position and frequency \citep[e.g.,][]{jp11}.
Such considerations may be used to optimize strategies for maximal survey yields.

\subsection{Radio frequency interference} \label{s:rfi}  

Impulsive and narrow-band RFI can be a major impediment in the detection 
of fast transients, increasing the number of false positives and raising the system noise. The issue of a false positive increase is particularly poignant for real-time detection schemes. 
With the ever-increasing number of (especially potentially astrophysically mimicking) RFI sources and the advent of wide-bandwidth 
observing systems,
it is becoming imperative to develop mitigation strategies for a wide variety of RFI sources and signals. 

Significant resilience to RFI can be developed through the use of appropriate 
instrumentation and online identification and excision schemes.
Systems that use multi-bit recording can have significant dynamic range advantage 
over the traditional one- or two-bit 
recorders used in most systems.  Prominent among prospective online mitigation schemes 
are those which employ median absolute deviation or spatial filtering 
\citep[e.g.][]{royetal2010,koczetal2010} and spectral kurtosis
filtering methods \citep{nita-gary2010}.  Effectiveness of a given strategy will depend
on the instrument as well as the RFI environment.

Even with online schemes, however, a large number of false positives may pass through the processing pipeline, requiring post-detection mitigation schemes so that the number of candidate events that require human scrutiny can be reduced to a manageable level. This is especially critical for systems that need to function in a commensal mode.
The long baselines of array instruments provide excellent capabilities here, 
enabling coincidence checks to allow identification and 
elimination of a large fraction of spurious events that are not common to all array 
elements. As outlined in \S\ref{s:tech}, this forms the key strategy for our 
transient detection scheme for the GMRT. Coincidence filtering provides a simple but powerful strategy.

Instruments with interferometric capabilities offer yet another powerful means of 
discriminating against RFI-generated transient events. The signatures of real signals 
are likely to be distinctly different in the image plane in comparison to those due 
to RFI. As such, by their very nature, short-duration transients may be originating 
from sources that are necessarily compact and hence will likely be seen as point 
sources in the image plane, provided an image can be made at sufficiently high time 
resolution \citep[e.g.,][]{law2012}. On the other hand, RFI bursts may yield various 
kinds of artifacts in 
the image plane, and are less likely to mimic the characteristics of point sources. 
Therefore by incorporating snap-shot imaging of candidate events among the event 
analysis strategies, further discrimination can be achieved against RFI sources. 


\section{The GMRT as a test bed instrument} \label{s:testbed}

The GMRT has a number of inherent design features which can be exploited for 
developing and demonstrating useful observing strategies for time domain science 
applications with next-generation instruments.
In addition to those previously noted, the combination 
of moderate-sized paraboloids and operation at low frequencies mean relatively large 
fields-of-view, e.g. $\sim$6 \sqdeg at 150 MHz, $\sim$1.5 \sqdeg at 325 MHz.
These,  along with the capabilities of its new software backend \citep{royetal2010}, 
in particular its ability to capture raw voltage data from all 30 array elements and 
make them available to software-based processing systems and pipelines, 
are promising for a variety of exploratory development. 

\subsection{The GMRT software backend} \label{s:gsb}
The recently developed GMRT software backend (GSB), built using mainly commercial,
off-the-shelf (COTS) components, 
is a fully real-time 32 antennas, 32 MHz, dual-polarization backend.  The basic 
requirements for the GSB are to support two main modes of operation : (i) a real-time 
correlator and beamformer for an array of 32 dual polarized signals with a maximum 
bandwidth of 32 MHz, (ii) a base-band recorder where raw voltage signals from all 
the antennas can be recorded to disks, accompanied by off-line correlation and 
beamforming.  Further details on design and implementation are described in 
\citet{royetal2010}. 


\begin{figure}[t]
\epsscale{0.9}
\plotone{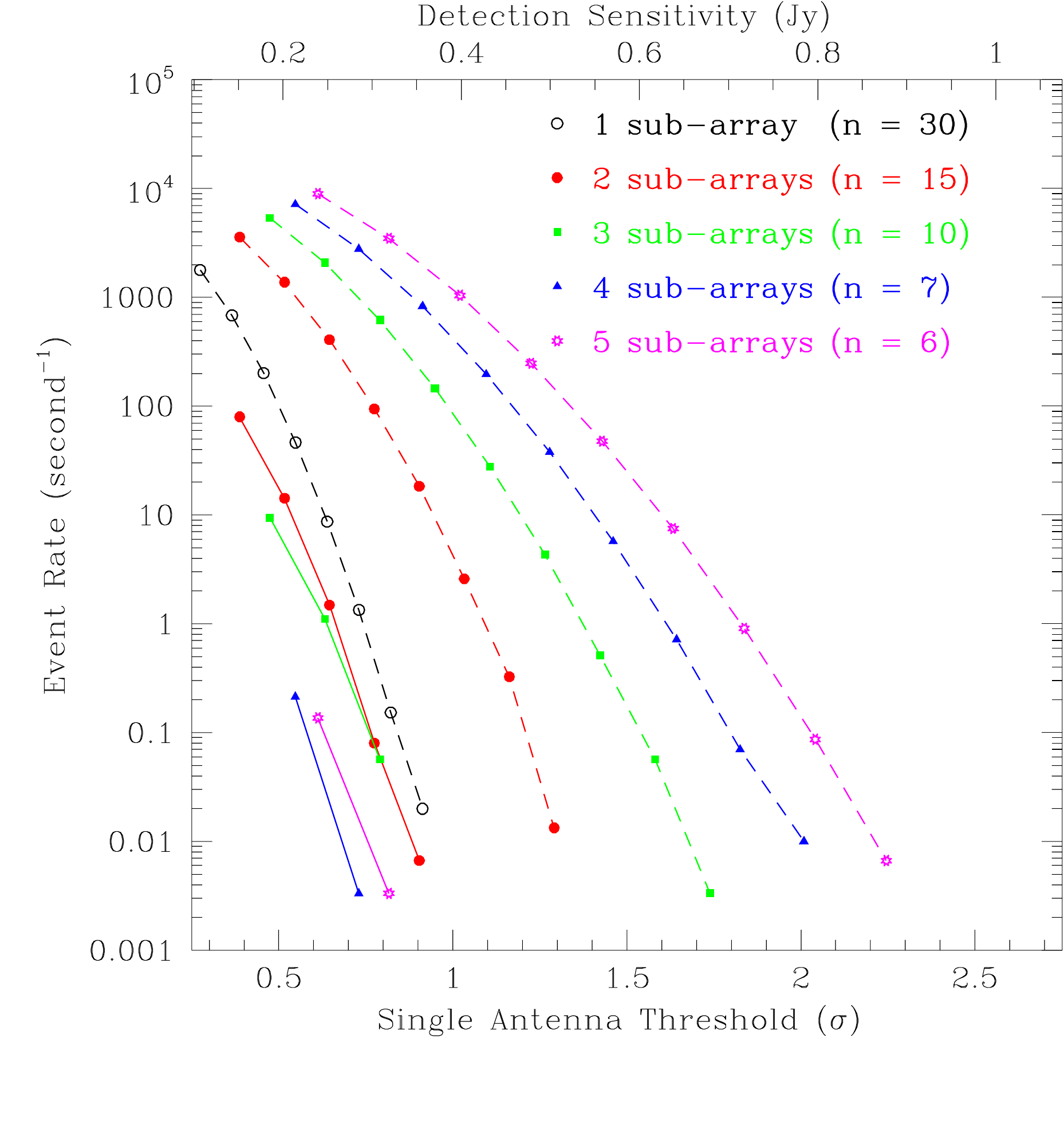}
\caption{Event rate vs. detection threshold for different sub-array combinations 
possible with the GMRT array, for a representative data set to characterize false 
positives due to signal statistics. For each combination, a pair of curves are shown, 
where the top and bottom (dashed and solid) ones correspond to the event rates at 
the pre- and post-coincidence filtering stage. 
The pre-coincidence (dashed) curves denote the aggregate event rates from multiple 
different sub-arrays (i.\,e.~the sum total of the event rates) for a given combination. 
Detection sensitivity is a strong function of the number of 
antennas  included in a given sub-array, while the coincidence power increases 
with the number of sub-arrays. }
\label{fig:dm200}
\end{figure}

\subsection{Transient exploration with the GMRT} \label{s:transient}

For transient exploration, our eventual goal is to develop and implement a 
system that will generate and process multiple incoherent array data streams 
in real-time for detecting transient candidate signals, and trigger the data 
recording system to extract and store relevant raw data segments for detailed 
offline investigations. Given the complexity of the problem, we adopt a 
two-phase strategy: the first phase involves conducting some pilot surveys 
and the development of a processing pipeline that operates on recorded raw 
voltage data. The outcomes from these are then used to finalise the design 
considerations for a real-time transient detection system.  Among the most 
powerful features of such an approach are:
\begin{itemize}
\item
exploitation of long baselines for powerful discrimination between signals of 
RFI  origin and those of celestial origin via effective coincidence filtering 
and cross-checks between multiple independent data streams. 
\item
event localisation possible via high-resolution imaging (5-10$^{"}$) and/or full 
beam synthesis across the FoV, both for important integrity checks as 
well as for facilitating high-frequency and multi-wavelength follow-ups with other 
instruments.
\item
the ability to form sensitive phased-array beams toward targets of interest and 
record baseband data so as to enable high time resolution studies of signal 
characteristics including coherent dedispersion and polarimetry. 
\end{itemize}

The modest recording bandwidth (maximum 32 MHz) of the current GMRT makes this a feasible 
exercise in terms of the related data rates and processing requirements. Even though 
the GMRT's FoV is relatively small in comparison to those of SKA pathfinder 
instruments such as ASKAP or MeerKAT, the gain of a single antenna of the GMRT is almost 
10 x  larger than that of a single ($d \sim 12$ m) element of these next-generation 
arrays. Thus, with an aggregate effective collecting area of $\sim$ 3\% SKA, the GMRT 
makes a highly sensitive instrument for conducting useful science demonstrations.  \\

\begin{figure*}
\epsscale{2.0}
\plottwo{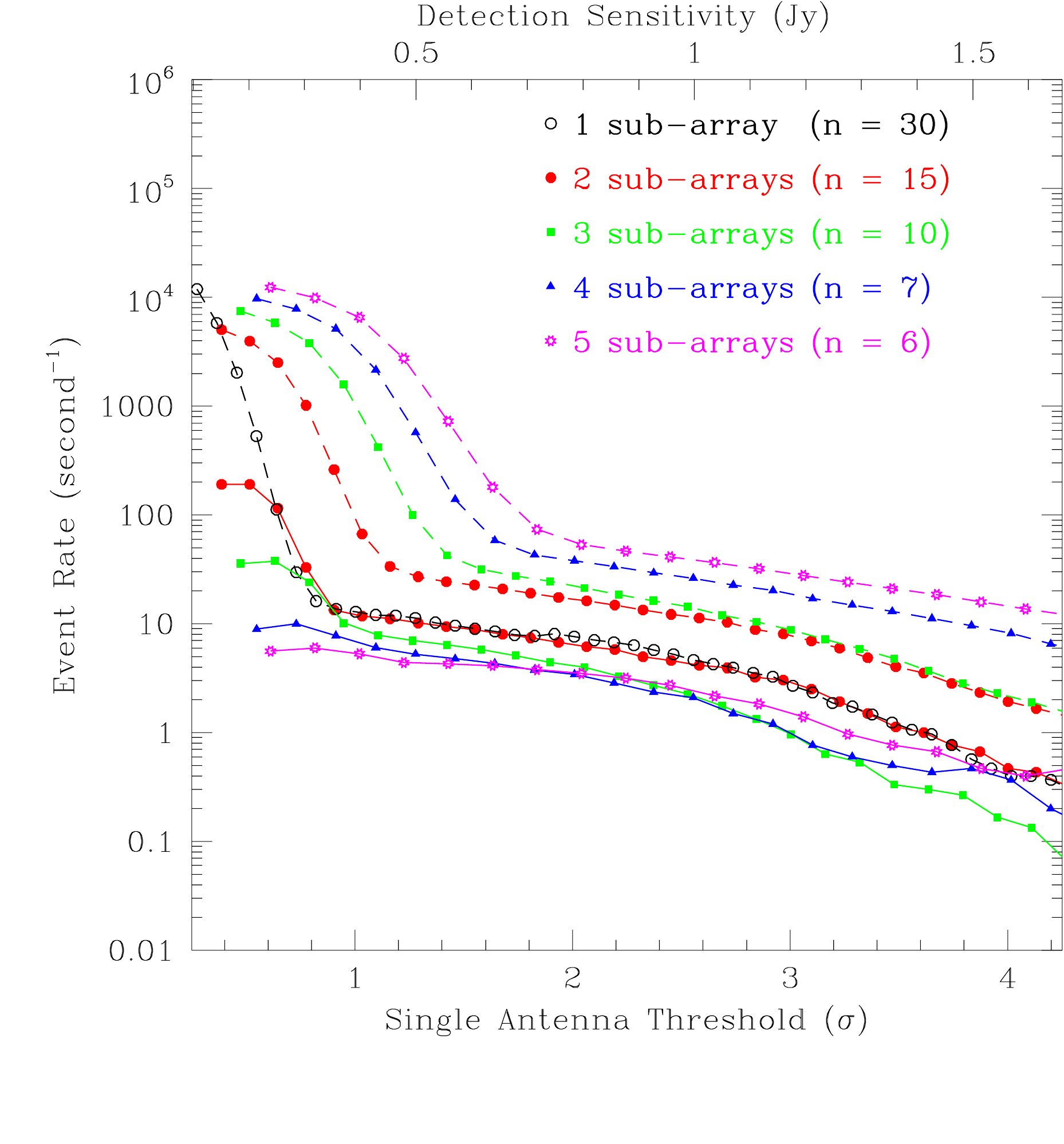}{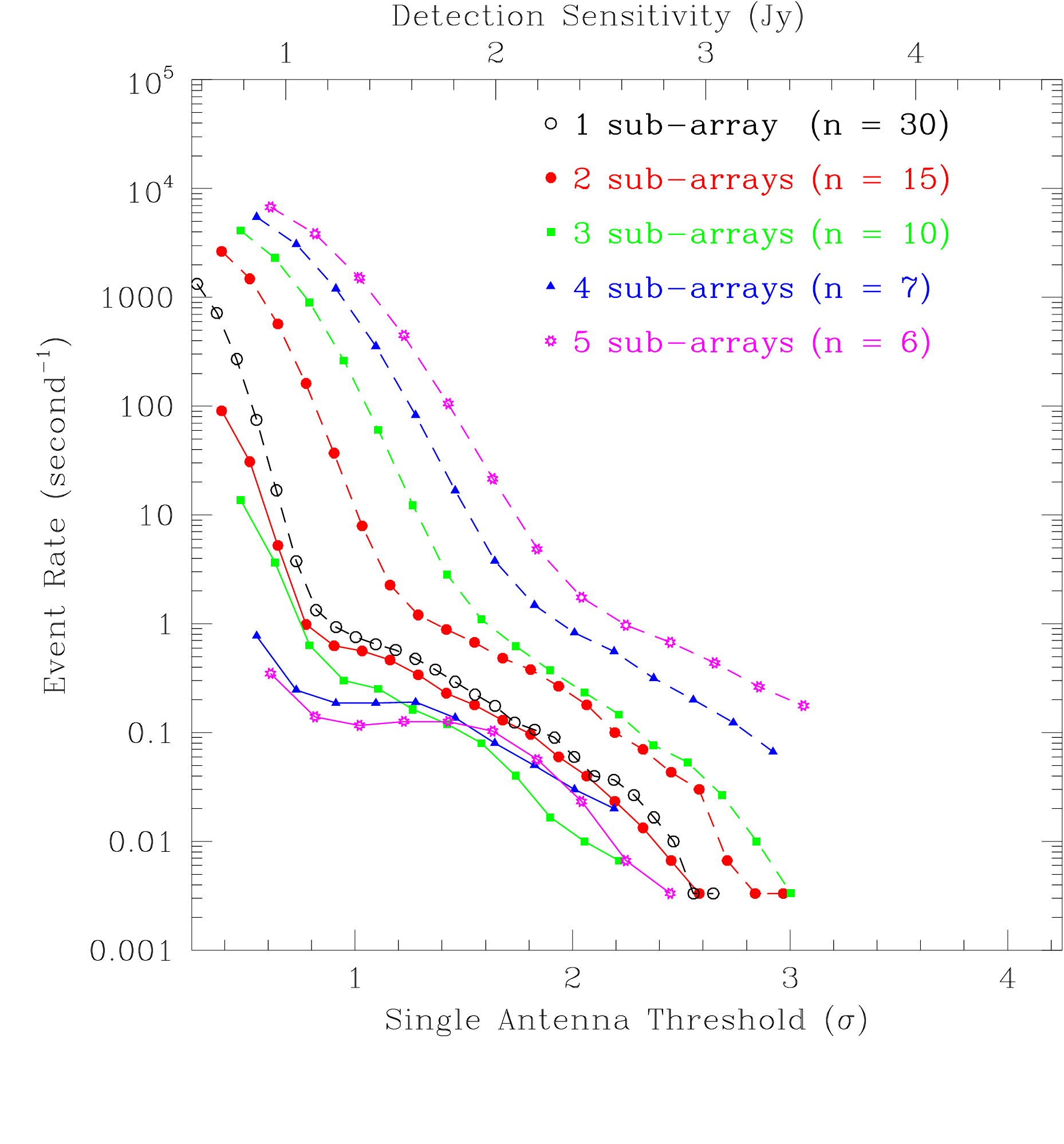}
\caption{Plots similar to Fig.~\ref{fig:dm200}, for a field encompassing a known pulsar. 
The pulsar (PSR J1752$-$2806) is located near the edge of the $\approx$ 1.5 ${\rm deg^2}$ 
FoV (i.\,e.~$\approx70^{\prime}$ offset from the phase center). 
{\it Left}:  processing at the pulsar DM and over the full recorder bandwidth 
($\Delta\nu$=16.66 MHz); {\it right}: processing over a narrow bandwidth 
($\Delta\nu/8$) to emulate weaker 
pulses. The detection sensitivity in Jy (top label) is based on nominal gain and 
system temperature for the GMRT at 325 MHz.} 
\label{fig:psrdm}
\end{figure*}

\subsection{A pilot transient survey with the GMRT} \label{s:survey}

In order to aid the related technical development and demonstrate the scientific 
credibility of transient exploration strategies, we conducted a pilot survey for 
short-duration transients with the GMRT, covering a small area of the sky  
($-10^{\circ} \le l \le 50^{\circ} $ and $ |b| \le 3^{\circ}$) with fairly short 
dwell times (5 minutes per pointing).  The data were collected in a specially
designed observing mode where raw voltage streams from all 30 antennas 
were recorded on to the disks. This survey was conducted at 325 and 610 MHz, 
where the GMRT offers its highest sensitivity, due to the large gains (G$\sim$10 \kpjy 
for the full array) and relatively low system temperatures (\Tsys $\sim$ 100 K) 
at these frequencies. 
The region within $1^{\circ}$ of the plane was surveyed at 610 MHz and the areas 
above and below this at 325 MHz. This choice was based on two main considerations: 
(1) to alleviate severe scattering at low frequencies, in particular very close 
to the plane and toward the Galactic centre; (2) to optimise the survey speed: 
${\rm 18~deg^2~per~hr}$ at 325 MHz vs ${\rm 5~deg^2~per~hr}$ at 610 MHz, so that 
the survey is completed within a modest amount of telescope time. The specific 
sky region was chosen because of its significant overlap with that of the Parkes 
Multibeam survey (thereby allowing immediate high frequency checks of any promising 
candidates), and also because it is the sky region where the density of known pulsars and
rotating transient objects is the largest. The relatively short dwell times mean that the survey 
is primarily sensitive to sources with fairly high event rates (10 ${\rm hr^{-1}}$ or 
more), such as giant-pulse emitters and rotating radio transients. 

In addition to the above pilot surveys, we also conducted observations in a number 
of exploratory modes. These include modes in which the array was sub-divided into 
multiple different groups (i.e. sub-arrays), with all configured to make pointed 
observations of a single selected target (such as the Crab pulsar), as well as modes 
in which different sub-arrays were configured to point to  different targets of choice 
(i.e. a variance of the Fly's Eye observing mode). These observations were made at 
a frequency of 610 MHz. 
 
Given our primary technical objective of developing a transient detection system and the 
required methodologies, it was imperative to record this survey data in the 
``raw dump'' mode of the GMRT software backend. This exploratory mode allows recording 
raw voltages from all 30 elements of the array, in two polarizations, with either 
2- or 4-bit digitization.  The aggregate data rate was approximately 
1 ${\rm GB \, s^{-1}}$ or 3.6 ${\rm TB \, hr^{-1}}$ (from 30 $\times$ 2 signal 
paths). For the survey parameters outlined above, this amounts to 42 hr of on-sky 
time, translating into a total data volume of 151 TB. These data were transported to 
the Swinburne supercomputing facility where all processing and analyses were carried 
out. Transient searches spanned up to 1000 \dmu in DM (in 1000 DM steps) and a 
maximum time scale of $\approx$500 ms. More details on the processing and results 
will be reported in a future publication. 

\begin{figure*}[t]
\epsscale{2.0}
\plottwo{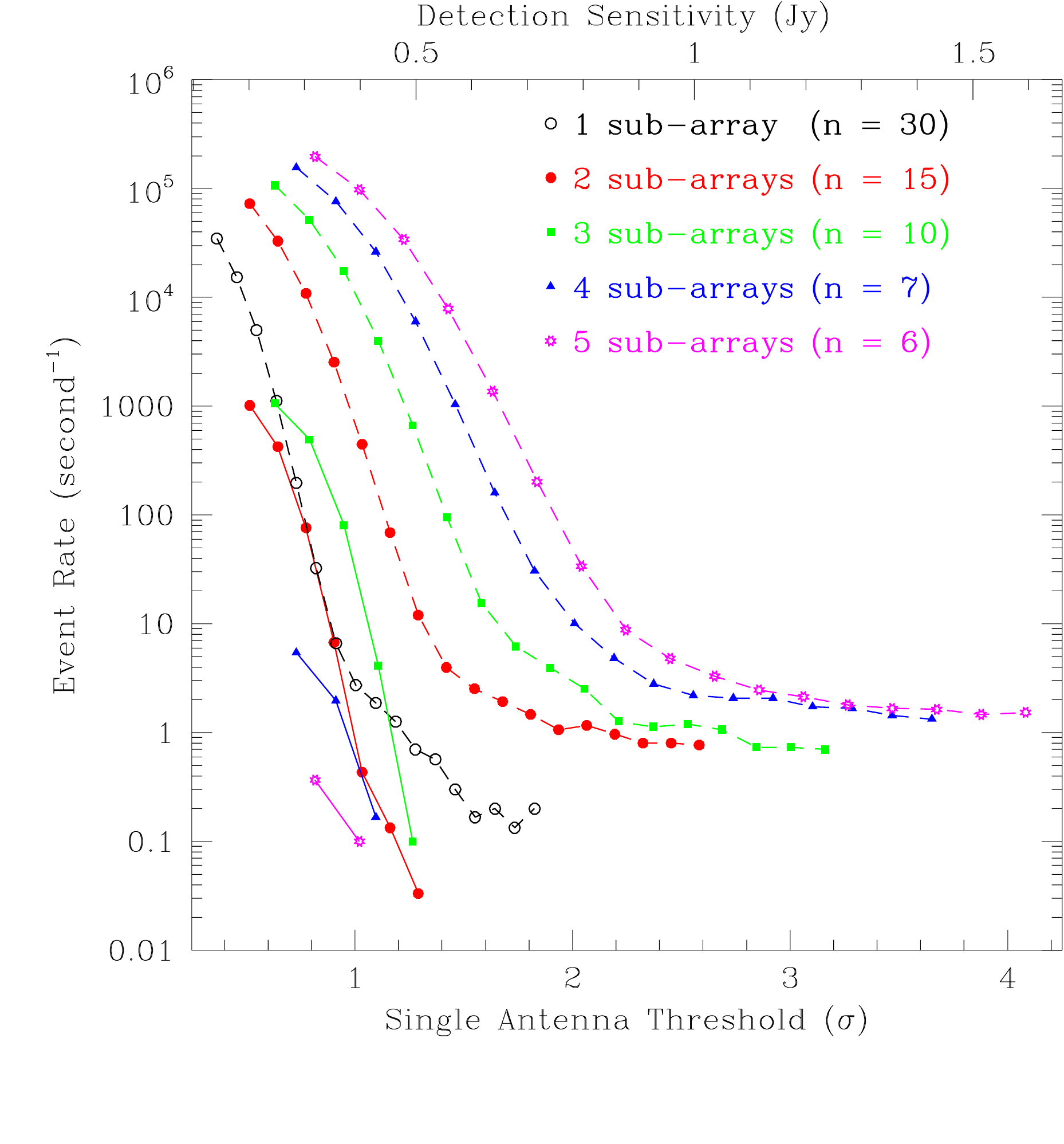}{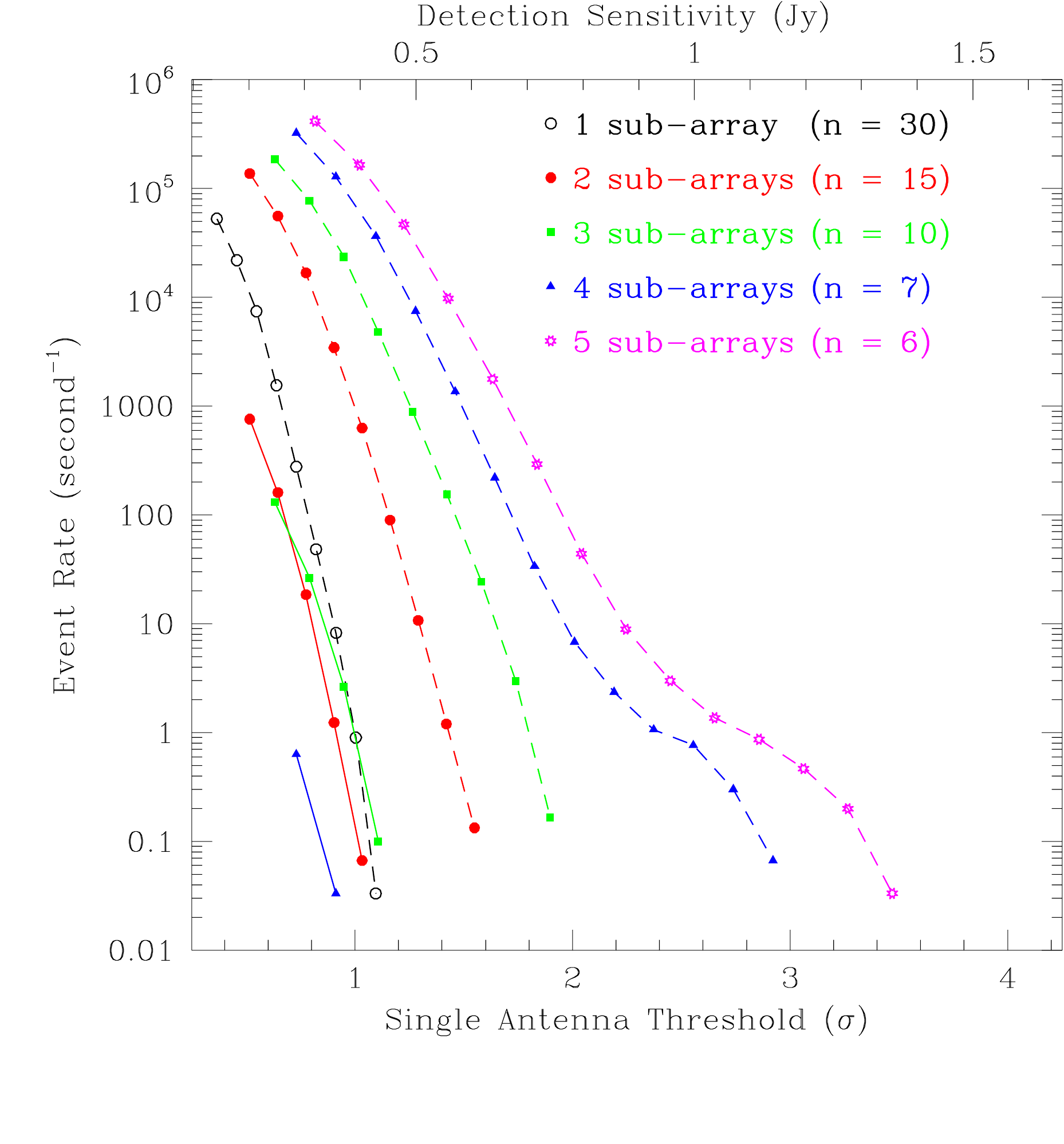}
\caption{Plots similar to Fig.~\ref{fig:dm200}, where the chosen fields correspond to 
``blank skies" (i.\,e.~fields that contain no detectable known pulsars or rotating radio 
transients) and during the conditions when the RFI-generated false positives were 
numerous. The left panel is the results for the survey field GTC\_352.17+1.27 when 
RFI was comparatively severe (in terms of the RFI-generated events), and the right 
panel shows the results for another field GTC\_002.25+1.17 when RFI was relatively 
modest.}
\label{fig:blanks}
\end{figure*}

\section{Transient detection pipeline} \label{s:pipeline}

In this section we outline a transient detection pipeline that we developed for 
offline analysis of the data from the pilot survey with the GMRT.  We delve into 
various steps involved as we proceed from raw voltage data to the detection and 
final scrutiny of candidate events.  The data from our pilot survey runs were 
used as test beds for developing the related software. In this paper we 
focus primarily on methodology and algorithms, with the implementation details to 
be reported in a separate paper. 

The basic idea involves generating multiple incoherent sub-array beams and using 
coincidence filter schemes for the rejection of false positives and RFI.  The array 
layout of the GMRT and a possible scheme for sub-dividing into multiple distinct groups 
(sub-arrays) is shown in Fig.~\ref{fig:layout}.  The block diagram of the processing
pipeline is depicted in Fig.~\ref{fig:pipeline}.  Even though our beamformer software 
has been heavily optimized for a specific architecture
(a constraint that arises from the GSB design considerations; see \citet{royetal2010}), 
the general scheme may 
also be applicable to other array instruments such as ASKAP or MeerKAT.

\subsection{Raw data capture from array elements} \label{s:rawdata}

As we demonstrate in later sections, access to raw voltage data from individual 
array elements offers a great deal of flexibility in terms of planning and conducting 
efficient transient searches with multi-element interferometric instruments. The 
raw voltage data can be easily interfaced to an incoherent beam former that offers 
the choice of the number of sub-arrays as well as the number of elements per sub-array. 
The sensitivity and other requirements of transient searching outlined in \S~\ref{s:tech} 
can thus be met with minimal constraints. Moreover, such flexibility can also be exploited 
to adapt to the changes in the RFI environment across the array. 

The baseband recorder mode of the GSB can be configured for either 16 or 32 MHz 
bandwidth, with 4 or 2 bit digitisation respectively, so that the aggregate data 
rate is limited to 1 ${\rm GB\,s^{-1}}$ or 3.6 ${\rm TB\,hr^{-1}}$ (again a 
constraint imposed by the GSB design considerations). The recording cluster used 
in the current system comprises 16 nodes, each with 4 TB of data storage, thereby 
providing a total data storage capacity of 64 TB, i.\,e.~a capability that can cater 
up to 18 hr of continuous baseband recording. There are four 1 TB disks connected 
to each node, and data from each antenna are streamed into separate disks.  Each 
recorded data buffer is accompanied with a timestamp derived from the NTP server. 

Online RFI detection and excision is an important consideration for transient 
detection with the GMRT. Of the prospective schemes described in \S~\ref{s:rfi}, 
filtering that relies on median absolute deviation is the only technique that has 
been tested on the GMRT data \citep[e.g.][]{royetal2010}. 
It is our aim to further explore the efficacies of this as well as other methods in the 
detection of short-duration transients, and converge on a possible implementation scheme for 
the real-time version of our pipeline. We have incorporated some rudimentary data 
quality checks in our current processing pipeline. These include basic sanity checks 
of each and every data stream for any instrumental failures or malfunctioning and 
then using this information to suitably reconfigure relevant sub-arrays  as well as 
the related coincidence parameters. 

\begin{figure*}
\epsscale{2.0}
\plotone{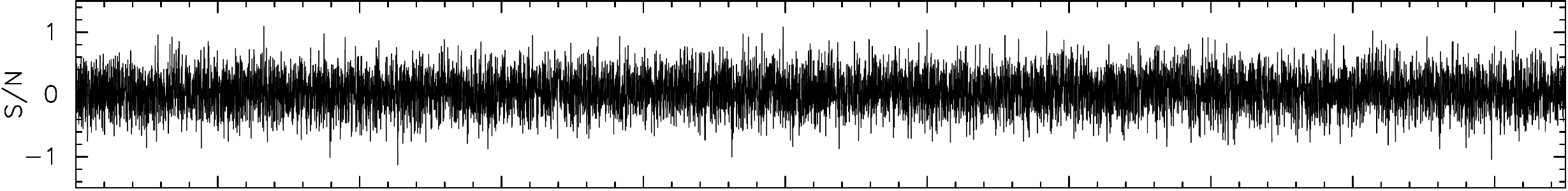}
\plotone{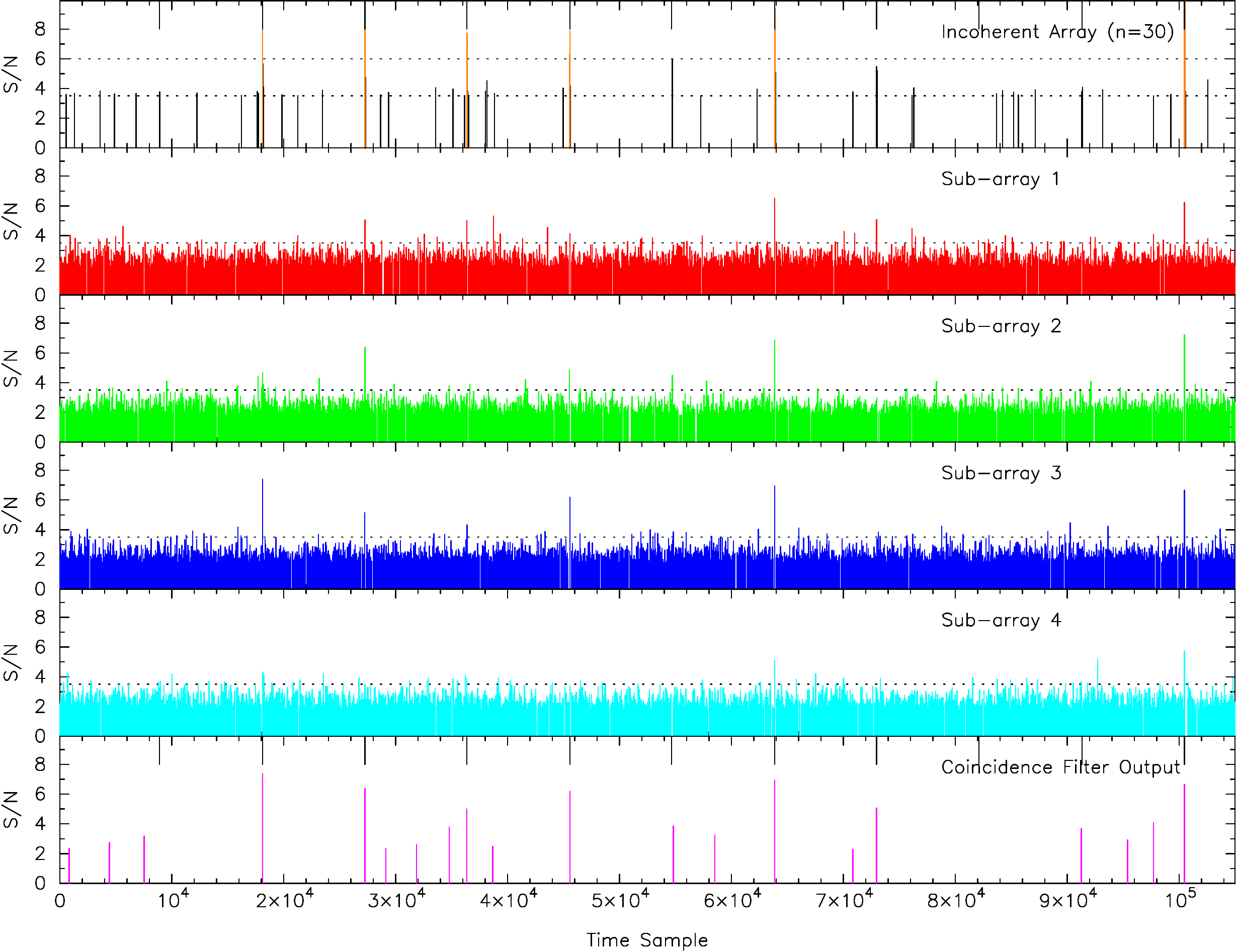}
\caption{An illustrative example to highlight the power of multiple sub-arrays and 
coincidence filter for the detection of real astronomical transient events whose 
peak amplitudes may be well below typically used 5-6$\sigma$ thresholds. Data from a 
field encompassing a known pulsar (but located at a large offset from the phase 
center) are processed over a small fraction of the recording bandwidth ($\Delta\nu/8$) 
to mimic such a scenario. The top panel shows the raw time series (at the original time
resolution of 30.72 $\mu$s), followed by the detections from a single incoherent 
sum of all 30 antennas for two set thresholds of 6$\sigma$ and 3.5$\sigma$ (i.e. second 
panel from the top); the four panels that follow (from third to sixth) show the detections (at 
a much lower 1.8$\sigma$) from individual 
sub-arrays when the array is sub-divided into four groups. The bottom panel is the output 
from the coincidence filter. The large tics in the second and bottom panels are the reference 
markers for the expected locations of pulses. 
Of the 11 pulses within this short data block (6 s), all but the faintest pulse have 
been successfully detected.} 
\label{fig:foursubs}
\end{figure*}

\subsection{Formation of multiple incoherent beams} \label{s:subarrays}

The rationale for dividing the array into multiple sub-arrays and opting for an incoherent 
addition of the intensities from telescopes in each sub-array is already detailed in 
\S~\ref{s:tech} (see also Fig.~\ref{fig:layout}). The simplest implementation of this 
procedure involves combining the signals from different elements of the array after 
the required delay and broad-band fringe phase corrections and spectral de-composition.  
This is currently realised through a software system that operates on 2 $\times$ \Nant 
raw data streams from the data acquisition cluster and 2 $\times$ \Nsub sets of 
antenna "masks" (where \Nsub is the number of sub-arrays), and performs the relevant 
signal additions in parallel. 
These multi-channel filterbank data streams, with a time resolution = 2 \Nchan /\fsamp, 
where \fsamp is the Nyquist sampling frequency and \Nchan the number of spectral 
channels, are then converted to intensities and summed after application of the 
suitable antenna masks.  These incoherently added intensity data are integrated (if 
needed) to achieve the desired time resolution. For example, for processing the data 
from pilot surveys, we have configured this incoherent  beamformer to output data 
streams with 256 spectral channels across the 16 MHz bandwidth ($\Delta f $ = 62.5 kHz) 
at a time resolution of 30.72 $\mu$s. 

Higher time resolution can only be achieved at the cost of a reduced spectral 
resolution. Searches at low frequencies inherently benefit from high resolutions 
in both time and frequency, and therefore this involves trade-offs in terms of 
the maximum DM that can be searched and the achievable time resolution. For example, 
a higher spectral resolution (i.\,e.~512-channel filter bank sampled at 61.44 $\mu$s) 
at 325 MHz will allow searching out to larger DMs while still not limiting the 
detectability of intrinsically narrow signals such as giant pulses, as they will be 
scatter broadened to $\sim$1 ms.  

\begin{figure*}[t]
\epsscale{2.0}
\plotone{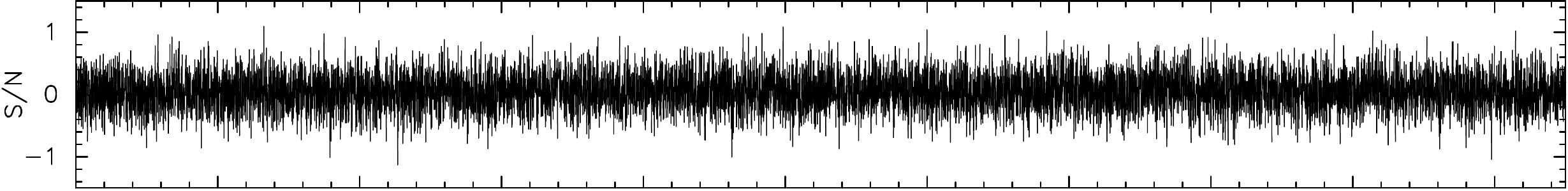}
\plotone{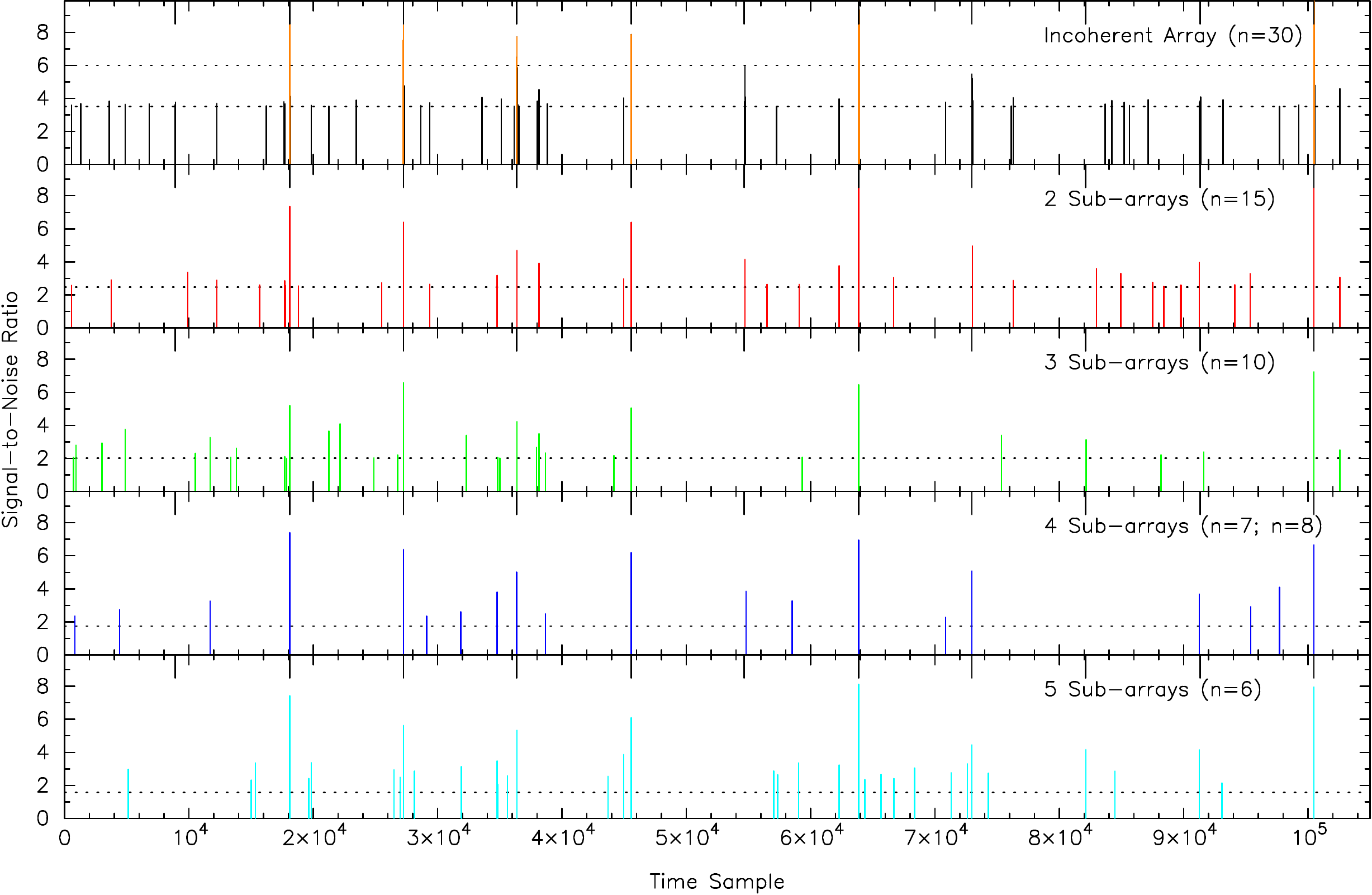}
\caption{Another illustration to highlight the advantages of multiple sub-arrays and 
coincidence for transient detection. The data block is the same as that was shown in 
Fig.~\ref{fig:foursubs} but processed for different possible sub-array groupings. The 
top panel shows the raw data time series, followed by the detections from a full 
incoherent 30-antenna array (i.e. second panel from the top), while the bottom 
four panels correspond to cases where the number of sub-arrays range from 2 to 5. The large 
tics in the bottom five panels are the reference markers that indicate the expected locations of
pulses.  The dotted lines correspond to the detection thresholds for each of the different 
sub-array combinations. The detection efficiency progressively improves from 2 to 4 
sub-arrays compared to a single incoherent sum, whereas 5 sub-arrays appears to be less 
than optimal choice. More quantitative analysis based on the processing of full data length 
(from this pointing) is summarized in Table 1.} 
\label{fig:allsubs}
\end{figure*}

\subsection{Searching for transients} \label{s:searching} 

\subsubsection{Dedispersion and Detection} 

Most traditional search algorithms for detecting fast transients operate on fast-sampled,
multi-channel (filterbank) data and hence involve incoherent dedispersion followed
by searching for transient events in the resultant time series.  Dedispersion is 
performed over a large number of trial DMs  (e.g. up to $\sim$1000 \dmu) using the 
standard direct dedispersion algorithm. This is the most computationally intensive part 
of our processing pipeline.  As dispersion delays can be substantial at the GMRT's  
frequencies (cf. \S~\ref{s:prop};  $\dtdm \propto \nu ^{-3}$ for small $\Delta \nu$), 
a large number of trial DMs are required to span such a large DM range, even when 
observations are made over moderate bandwidths of 16-32 MHz.  

Our current processing pipeline makes use of the dedispersion software that was developed 
for the ongoing high time resolution survey at Parkes \citep{keithetal2010,bsetal2011}. 
This software takes advantage of modern multi-core processors that allow multi-threaded 
software to achieve significant speed-ups in computation.  While the Parkes survey decimates 
the data to two bits per sample, our processing pipeline is designed to operate on 16-bit 
data samples which provides a much higher dynamic range, and also ensures immunity against 
possible signal saturation from powerful RFI bursts. Following the convention in pulsar 
searches, spacings between DM values are determined by an analytic constraint on the 
signal-smearing due to incorrect trial DM. A GPU implementation of this dedispersion 
software has recently been developed \citep{barsdelletal2010} and is being integrated 
into the real-time version of our processing pipeline which will be described in a 
forthcoming paper. 

Our approach of employing multiple sub-arrays for transient detection with the GMRT means 
that the dedispersion stage will result in $ \Nsub \times \Ndm $ time series to search 
for, where  $\Ndm$ is the total number of DM trials. As we demonstrate in later sections, 
four sub-arrays are optimal for transient detection with the GMRT. For our observing 
frequencies and recording bandwidth, ensuring a signal degradation (from dispersive 
smearing due to incorrect DM) of no more than 1\% will necessitate $10^3 $ time series 
for each sub-array. 

Detection of transient events essentially involves the identification of data samples, 
or groups of samples, that are above a set threshold (e.g. 5$\sigma$) in the dedispersed 
time series.  Matched filtering, as approximated through a range of box car widths, is 
the commonly employed detection technique \citep[e.g.][]{cm03}. This simple, yet 
effective, methodology has been extensively used in a number of ongoing searches 
based at Parkes and Arecibo as well as other instruments around the world 
\citep[e.g.][]{denevaetal2009,bsetal2011,bhatetal2011}. Alternate techniques, such as 
those based on quadratic discriminants and other statistics, have also being explored and 
demonstrated to a certain extent \cite[e.g.][]{thompsonetal2011,fridman2010,spitleretal2011}, 
though their efficacies as viable alternatives in large-scale transient searches have not 
yet been thoroughly tested.  The present version of our pipeline therefore employs matched 
filtering as the primary detection algorithm.

\begin{figure}[t]
\epsscale{0.77}
\plotone{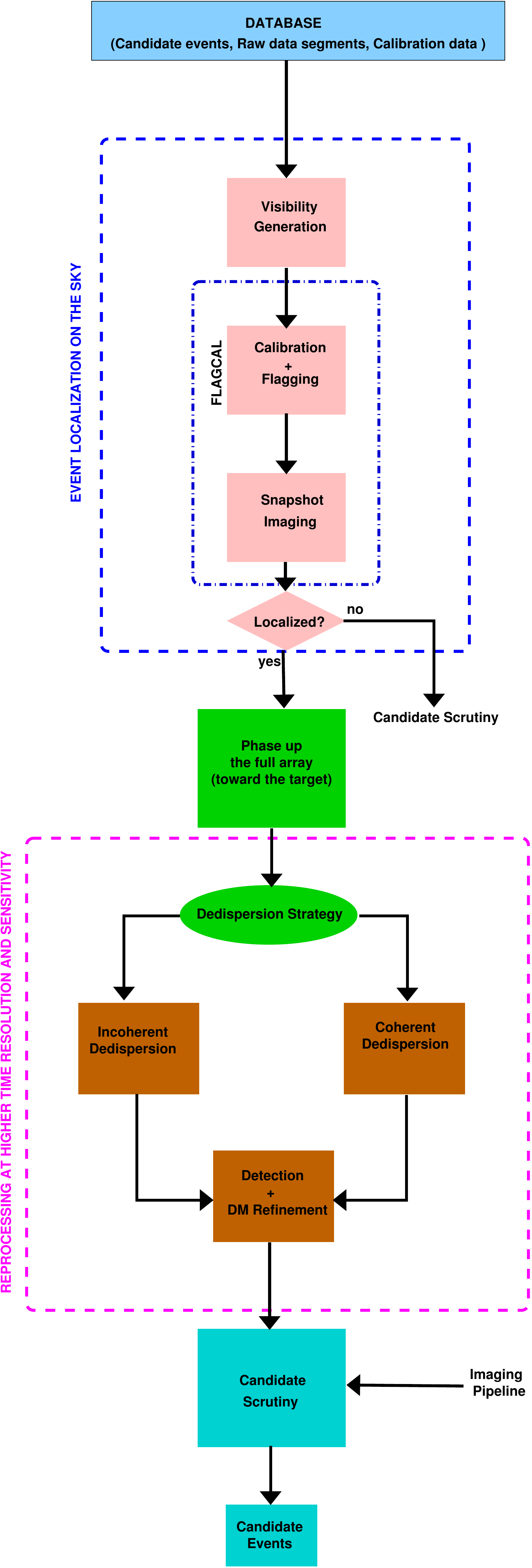}
\caption{
Diagrammatic representation of the event analysis pipeline for a detailed scrutiny 
of the candidate events that emerge from the GMRT transient detection pipeline. 
Depending on the characteristics of the candidate event, one or more analyses
or scrutinies may be possible, e.g. processing of the raw data segments to generate 
the visibilities (for making a snapshot image of the FoV); forming a sensitive 
phased-array beam toward the target (to enable a high quality detection), followed 
by a suitable dispersion removal process and the signal detection. Final event 
scrutiny is then performed to arrive at a list of candidates for further follow-ups. 
} 
\label{fig:eventpipe}
\end{figure}

Matched filtering is approximated by progressive smoothing of time series data 
over a range of box car filters of widths $2^n$ samples, followed by application 
of threshold tests, each time recording the event amplitude, the time of occurrence and duration 
\citep[e.g.][]{denevaetal2009,bsetal2011}. While relatively simple and easy to implement, 
this technique of matched filtering  has some shortcomings; for instance, 
powerful RFI bursts that occur over relatively long durations  (i.\,e.~\Wp $\gg$ $dt$) 
will be detected as an overwhelmingly large number of events. Furthermore, as the pulse 
templates have discrete widths of $2^n$ samples by design of the algorithm, this means 
reduced sensitivity to events whose widths are intermediate to those of the chosen box 
car filters. As discussed by \citet{denevaetal2009} and \citet{bhatetal2011}, alternate 
methods, such as those based on time domain clustering along the lines of the 
{\it friends-of-friends} logic may help alleviate some of these demerits of matched 
filtering.

\subsubsection{Initial scrutiny of detected events} \label{s:initial}

An important consequence of the above described search strategy, which involves many 
trial DMs and box car widths, is that each single event will be detected as multiple 
events in the DM--time parameter space depending on the signal strength, duration and 
DM. In order to identify and associate multiple related events arising from a certain 
transient pulse, we employ algorithms along the lines of {\it friends-of-friends} logic 
that is very similar to those described in \citet{bsetal2011}. This essentially involves 
performing an {\it association} of events in time, DM and the matched filter width \Wp, 
effectively identifying the groups of related events in the parameter space. In practice, 
this may be realized in two steps; first, for a given DM, association of related events 
is performed in time; specifically starting with the widest pulse and looking for pulses 
which overlap and, as each new pulse is associated, the search window is extended to 
include the net time range. The criterion here is that the peak of the second pulse 
overlaps with the full width of the first.  A similar association is subsequently 
performed in DM, by effectively checking for {\it contiguous} events in DM and 
associating any events with the S/N characteristics that may follow likely astronomical 
signals (e.g. peaking at true DM and lower S/N with an increase in departure from the 
true DM).  The procedure also accounts for possible time delays or advances expected due 
to the DM offset, thus ensuring that multiple real events of different DMs are detected as 
separate events, whereas multiple events due to a given RFI burst (often spanning the 
full DM range) are still detected as a single event. In the end, multiple points in the 
DM-time-\Wp parameter space which are related are counted as a single event.

\subsection{Coincidence filtering and elimination of false positives} \label{s:coin} 

The main goal of coincidence filtering is the removal of false positives due to noise 
and RFI, thereby improving the efficiency to discriminate genuine signals of astronomical 
origin. In ideal conditions, when the signals are sufficiently strong to allow clear 
detections, this can be achieved with strict simultaneity checks in terms of the 
characteristics of the detected events.  An example detection of this kind is shown 
in Fig.~\ref{fig:example}. However, real-world considerations necessitate 
a more flexible approach, especially when the detected signals are relatively weak 
(i.\,e.~near the detection thresholds) and the sensitivity of the sub-arrays is not 
guaranteed to be identical. 
An example of these kind of effects can be seen in  Fig.~\ref{fig:foursubs}, which shows
the detections from 4 sub-arrays of the GMRT, for a few successive pulses of a relatively
weak pulsar where the detections are barely above the acceptable S/N threshold.  As can
be seen, sub-array 4 has a somewhat lower sensitivity than the other 3 sub-arrays, and
even otherwise, the detectability of individual pulses does vary across the sub-arrays, 
presumably due to noise fluctuations.
In such cases, it is possible that a given transient pulse may be detected at slightly 
different DMs, pulse widths (durations) or times of occurrence by different sub-arrays
(an example for which is shown in Fig.~\ref{fig:weakpulse}), 
and an efficient recovery mechanism needs to take this into account. 

In order to account for such effects and their potential impact on the detectability 
of genuine astrophysical signals, we have devised a somewhat flexible coincidence logic 
for the GMRT transient pipeline. Basically, each event is characterized by its basic 
properties such as the arrival time ($t$), duration ($w$), DM and the peak S/N. 
The events from different sub-arrays are cross-checked for coincidence criteria 
defined in terms of these parameters. Two events $ E_1 ( t_1 , w_1 , DM_1 , s/n _1 ) $ 
and $ E_2 ( t_2 , w_2 , DM_1 , s/n _2 ) $ from the sub-arrays 1 and 2 are treated to 
be coincident provided (i) overlap in their times of occurrence is within a set range, 
$T _{tol}$; (ii) the difference in DMs is within a set range, $DM _{tol}$; and 
(iii) the difference in the peak S/Ns is within a set range $S/N _{tol}$. That is, 
their characteristics need to be such that (i) $ \delta DM \le DM _{tol} $, 
(ii) $\delta (S/N) \le (S/N) _{tol} $, (iii) $\Delta T / w_1 \ge T _{tol}$ and 
(iv) $ \Delta T / w_2 \ge T _{tol}$, where $\delta DM$ and $\delta (S/N)$ are the 
fractional differences in the DMs and S/Ns, and $\Delta T$ is the overlap in the 
time ranges, as determined by the respective time ranges, $ (t_1 , t_1 + w_1 )$ and 
$ ( t_2 , t_2 + w_2 )$ for the events $E_1$ and $E_2$. For coincidence between three 
or more sub-arrays, each event is checked against events from every other sub-arrays, 
beginning with the highest S/N event. As we illustrate in later 
sections, such a scheme is particularly important for the detection of weaker signals.
Specifically, it increases the prospects of detecting 
real (weaker) signals, while limiting the number of false positives. 

The parameters $T _{tol}$, $DM _{tol}$ and $ S/N _{tol} $ thus determine the 
"stringency" of the coincidence logic. For example, insisting for higher time 
overlaps (e.g. $T_{tol}$ \ga 50\%) between the detected events imply a stringent 
coincidence logic, whereas allowing for smaller time overlaps (e.g. $T_{tol}$ \la 20\%) 
would result in a coincidence that is relatively more lenient. The parameters 
$ DM _{tol} $ and $ S/N _{tol} $ essentially help ensure that only those events of 
similar characteristics (in DM and brightness) have chances of passing through the 
coincidence filter. The choice of the above parameters also has implications in 
terms of the rates of false positives, since a more lenient coincidence logic means 
relatively higher rates of false positives. Similarly, a higher stringency in the 
coincidence logic may potentially result in filtering out real astronomical signals 
that are near the detection thresholds. Events that pass the set coincidence criteria 
are subjected to a further detailed scrutiny while those that fail are rejected 
from the analysis.  While there may be various factors that influence optimal values 
of these parameters, RFI can also be expected to play a significant role, in particular 
for the GMRT given its location in a relatively RFI-prone environment. 

On the basis of a preliminary analysis of our survey data (i.\,e.~130 fields covering a 
$\sim$200 ${\rm deg^2}$ of the sky) at 325 MHz, it appears that representative values 
may be $\sim$5--10\% for $DM_{tol}$, and $\sim$50\% for $(S/N)_{tol}$ and $T_{tol}$. 
Specifically, we note that these are derived particularly from the data on two specific 
survey fields (GTC\_002.52--1.64 and GTC\_001.01--1.43) that contained  a known pulsar, 
but at relatively large offsets of 71' and 42' respectively from the beam phase center 
(i.\,e.~near the edge and well outside the half power beam). These were processed for 
various possible combinations of the parameters, and the tolerance settings that resulted 
in maximal number of real pulse detections (and minimal number of false positives) 
were treated as optimal choices. These were subsequently verified using data from the 
pointings at the beginning of the survey observations (when the pulsar would be at the 
phase centre).

\subsection{Examples of transient detections} \label{s:examples}

Figure~\ref{fig:example} shows an example candidate event detected in our survey 
observations (GTC\_001.01--1.43) that contained a known pulsar at an offset of 
1.2 deg from the phase center. This is from observations made at 325 MHz (i.\,e.~a 
FoV $\approx$1.5 \sqdeg or a half power beam width $\sim84^{\prime}$). 
These basic diagnostic plots illustrate a number of signal characteristics expected 
of astrophysical signals. For instance, the dedispersed time series and frequency-time 
plots (top panels) provide immediate assessments of coincidence of signal detection 
in multiple different sub-arrays. Other important signatures include a dispersion 
sweep in the time-frequency plane and the change in signal strength versus DM, which 
is shown as the dedispersed time series at the candidate DM as well as at two nearby 
DMs along with that at DM=0 (bottom panels). Our processing pipeline also records 
additional information such as signal strength vs. DM (for optimum \Wp) and signal 
strength vs Wp (for optimal DM). A basic scrutiny along these lines can be employed 
in order to arrive at a list of candidates that may require further detailed investigations.

\begin{figure*}[t]
\epsscale{2.25}
\plottwo{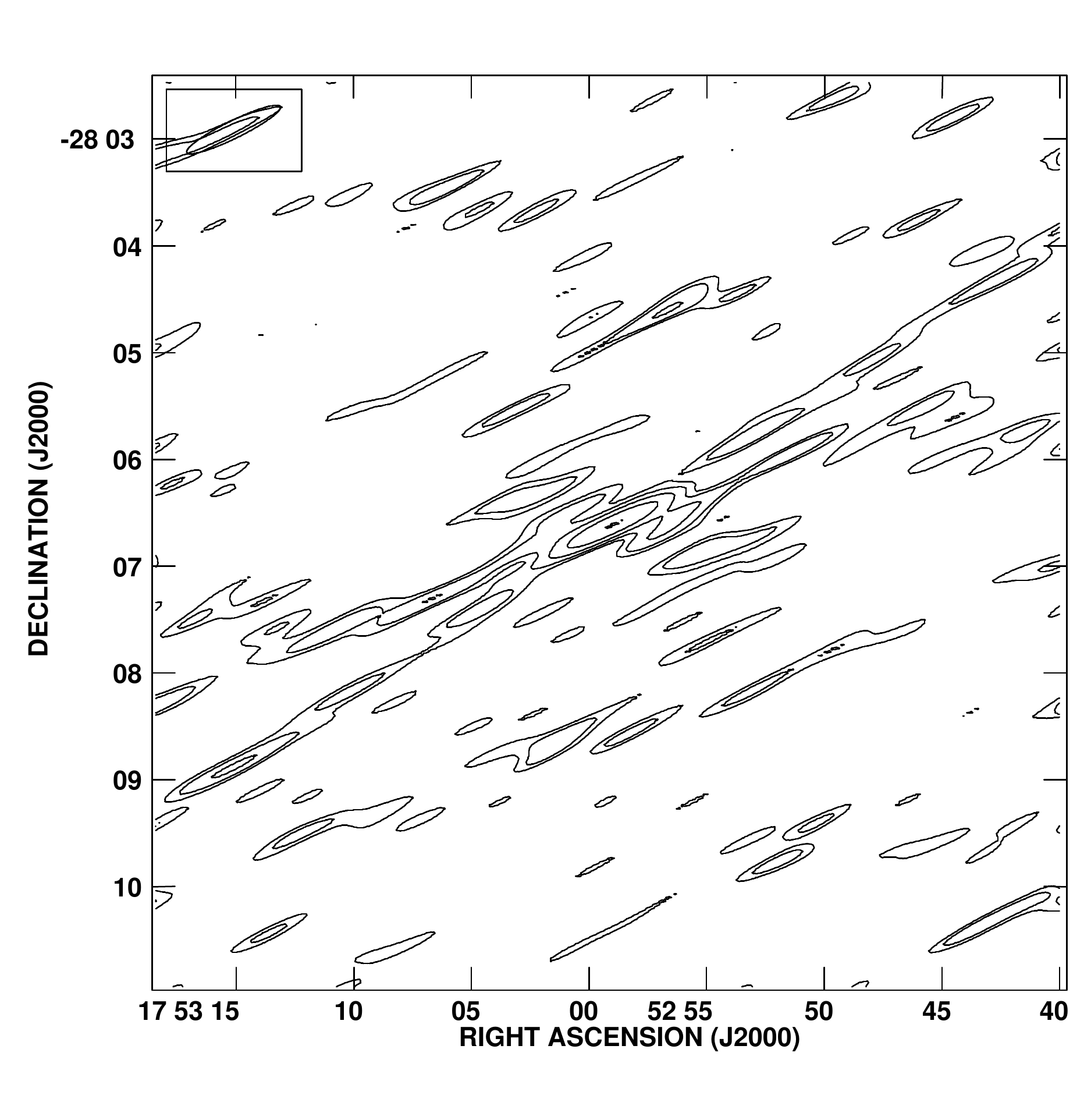}{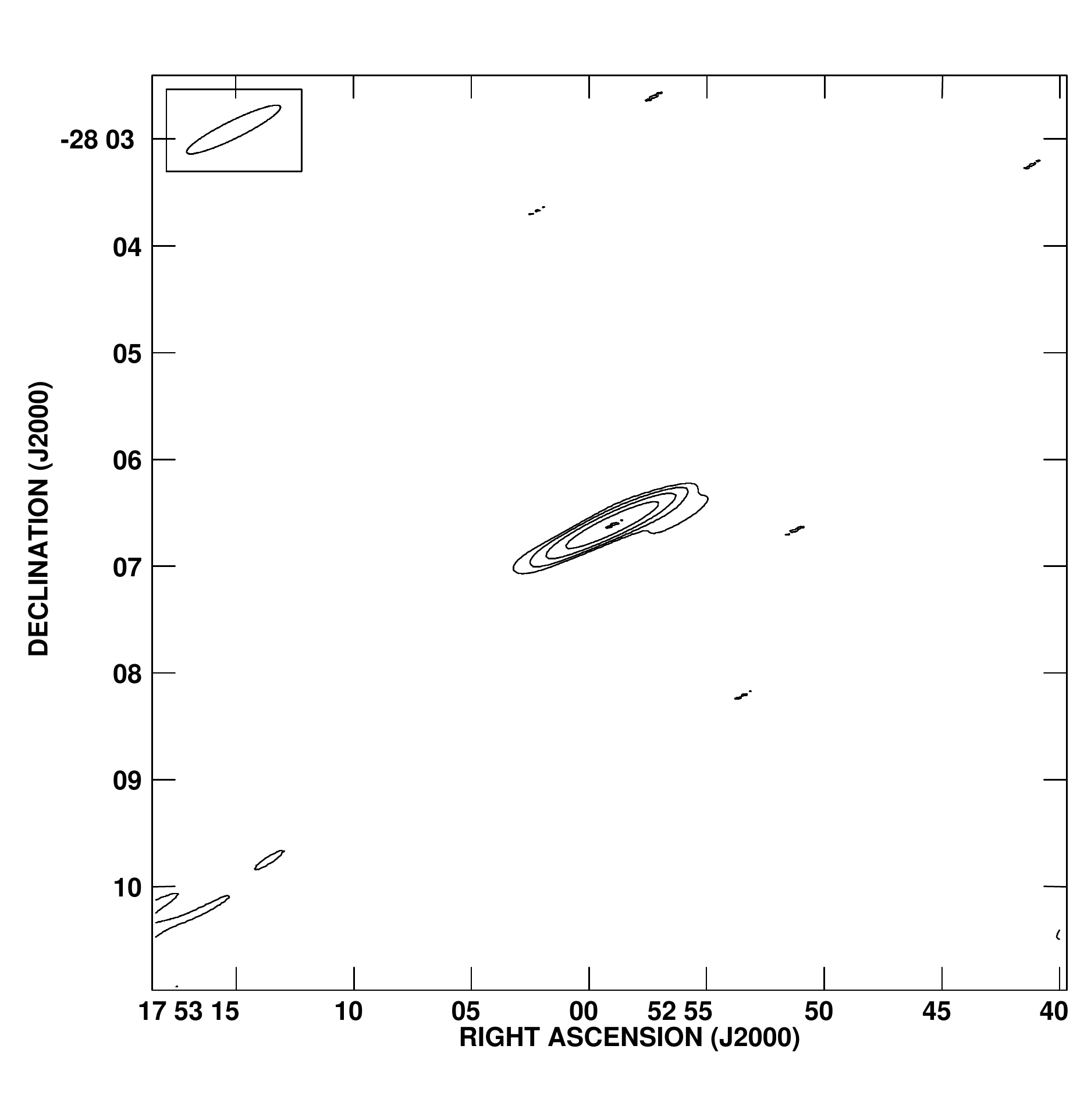}
\caption{ Images of a single pulse from the pulsar  J1752$-$2806. The left panel shows 
the ``dirty'' image, the sidelobe response of the interferometer can be clearly seen. The 
right panel shows the image after cleaning. The contour levels are logarithmically spaced 
between 20 and 270 mJy. The synthesized beam is shown in the top left corner and is 
$59^{''} \times 10 ^{''}$ in size. The flagging and calibration was done using the \flagcal 
pipeline, and the image was made using the AIPS task IMAGR.}
\label{fig:1748image}
\end{figure*}

Fig.~\ref{fig:crabgiant} shows another example from our transient detection pipeline. These 
observations were made in a special mode where seven of the 30 antennas were pointed to the 
Crab pulsar, thus emulating 7 sub-arrays, each comprising a single antenna. This provides a 
powerful coincidence filter against spurious events of RFI origin. The data were collected in 
a `survey mode', by making the telescopes scan the sky region around the Crab pulsar at a
rate of 0.5 ${\rm deg~min^{-1}}$. A bright giant pulse was thus detected as a `transient' 
when the pulsar was within the telescope beam (the half power beam width at 610 MHz is 
$\sim$0.5$^{\circ}$). This example highlights the need to employ very short DM spacings 
as well as high time and frequency resolutions in order to retain sensitivity out to durations 
as short as tens of $\mu$s, which is possible with our transient pipeline.

\section{Applications to Real Data} \label{s:eventrates}

In \S~\ref{s:tech} we discussed the advantages of using multiple sub-arrays and 
coincidence for transient detection and its impact on the detection sensitivity to 
transient signals. In \S~\ref{s:coin} we delved into the details of 
practical implementation of our coincidence detection logic. In this section we 
present some examples to illustrate the effectiveness of such a scheme through 
its applications to real data obtained from our pilot surveys. Specifically, we 
highlight (i) the reduction of false positives for the cases of (a) pure noise, 
and (b) RFI contamination (\S 5.1), and (ii) the power of coincidence filtering 
in facilitating the detection of weaker astronomical signals (\S 5.2).  

\subsection{Reduction of false positives} 

The basic underlying concept here is that the vast majority of events generated 
due to noise fluctations and RFI signals will be uncorrelated between different 
sub-arrays. In order to investigate this, we compare the pre- and post-coincidence 
filtering event rates as a function of number of sub-arrays and detection threshold. 
We consider various possible combinations for the GMRT array, ranging from 2 to 5 
sub-arrays, and process the data over a wide range of detection thresholds, down 
to $\sim$0.5$\sigma$ at the single antenna level.  This analysis becomes fairly 
cumbersome given the complexity of the number of sub-array combinations and running 
the pipeline at the very low threshold values that we use.  We therefore focus on some 
select case studies as described below.
 
\subsubsection{Case study 1: A blank field in the absence of RFI}

This is probably the simplest case, the results from which can be directly compared 
against the predictions based on the theoretical analysis presented in \S 2.1. To 
emulate an ``absence of RFI'', we performed the related analysis on a data set that 
is virtually devoid of any noticeable RFI.  To further reduce the effect of interference 
signals and keep matters simple, the data were processed at a single, large DM value 
(200 \dmu). Furthermore, the coincidence logic described 
earlier in \S~\ref{s:coin} was heavily simplified in order to match the assumptions 
made in the theoretical analysis (for instance, a uniform detection sensitivity 
across all different sub-arrays). Specifically, the tolerances in terms of DMs, 
arrival times and S/N ratios were set to zero, which implies the maximum possible 
stringency achievable with the coincidence logic. 

The results from the analysis are shown in Fig.~\ref{fig:dm200}, where we plot both 
the pre- and post- coincidence filtering event rates for detection thresholds down to 
$\sim$0.5$\sigma$. 
The sub-array groupings of antennas have been chosen such that maximal resilience against localized 
RFI is achievable; e.g. in the case of four sub-arrays, 3 of them are formed from 7-8 
antennas that are located along the east, west and south arms, whereas the fourth one is 
comprised of antennas from the central 1 km x 1 km area. 
The results for different sub-arrays have been scaled to equivalent single antenna 
thresholds by applying the theoretically expected scaling.\footnote{Under the assumption of identical gains and system temperatures for 
individual antennas, the detection threshold for a sub-array of $\nant$ antennas, 
$\sigma _{\rm sub} = \sigma / \sqrt{\nant}$, where $\sigma$ denotes the detection 
threshold for a single antenna.}
The top panel denotes the thresholds in units of Jy, assuming nominal 
sensitivity parameters of the GMRT (cf. Eqn.~\ref{eq:sens}). 
 The 10 to 100 times improvement seen in the post-coincidence 
event rates compared to a single 30-antenna sub-array (i.\,e.~the full GMRT array) 
is in rough agreement with the theoretical predictions (cf. Fig.~\ref{fig:jayanti}, 
where the region of interest are the first 5 curves in the top left hand corner of 
the figure).  There are however some discrepancies; for instance, the results for the 
3 sub-array case are only marginally better compared to those for 2 sub-arrays. Moreover, 
little improvement is seen in going from 4 to 5 sub-arrays. These discrepancies may be 
due to some faint RFI signals that are common to different sub-arrays, or perhaps 
because of detection thresholds not scaling as theoretically expected in practice.
 Besides this, the results are in accordance with expectations from the theoretical 
 analysis, thereby ratifying our basic principle for splitting the array into 
 incoherent sub-arrays.

\subsubsection{Case study 2: A blank field in the presence of RFI}
 
The GMRT's proximity to densely populated regions and operation at low frequencies 
pose major challenges in terms of corruption due to RFI. The main sources of RFI 
include power lines, transmitters, TV boosters and cell phone towers.  Some of these
are highlighted in \citep{pacigaetal2011}, which presents a detailed characterization 
of RFI sources seen in the GMRT's $\sim$150 MHz band.  A somewhat similar situation 
prevails at the other low frequency bands of the GMRT, even though it is true that 
most of the wideband, impulsive RFI sources have spectra that become weaker at higher 
frequencies.  The preliminary processing of our pilot survey data suggests that at 
least some modest fraction of our survey data at 325 MHz is corrupted by RFI.
The presence of multiple bright RFI sources, and short to moderate baselines of 
the array ($\sim$100 m to $\sim$25 km), lead to some interesting challenges in 
terms of gaining immunity against resultant false positives.  

Different survey scans with varying degrees of RFI were identified and analyzed 
in order to investigate the improvement in terms of the pre- and post-coincidence 
event rates. One specific example is shown in Fig.~\ref{fig:blanks}. 
The combinations of 4 or 5 sub-arrays retain the discriminatory power in terms of 
immunity from RFI false positives, resulting in a significant improvement in terms 
of the rates of false positives compared to the reference case of a full incoherent 
array.  However, the results for 2 and 3 sub-arrays are somewhat puzzling, 
particularly with regard to the improvement seen in going from 2 to 3 sub-arrays. 
In fact the 3 sub-array combination seems to be virtually ineffective when RFI gets 
severe (see left panel). This may suggest correlated RFI events between 3 sub-arrays; 
e.g. powerful RFI sources located near the central square region, the antennas from 
which are roughly evenly split between the 3 sub-arrays. While we have attempted to 
further investigate this puzzling observation by trialing different ways to form 
the sub-arrays (e.g. based on the proximity to the central electronics) and also 
by processing several different data sets, the results have not been quite conclusive. 
We therefore speculate this is likely to be an intrinsic feature of the GMRT array. \\

\subsection{Detection of real astronomical signals}

As outlined in \S~\ref{s:tech}, an improved efficiency in terms of the rates of 
false positives can only be achieved at the cost of reduced sensitivities at 
individual sub-array levels. While our approach to divide the array into multiple 
groups of incoherent sums seems like a reasonable trade-off, the net result is 
reduced detection sensitivities, particularly to the detection of weaker signals. 
For instance, a 6$\sigma$ transient pulse from a single 30-antenna array will be 
detected as a 3$\sigma$ event when the array is sub-divided into four groups. An 
equally important aspect therefore is the efficiency that may be achievable in the 
detection of such weak (but real) astronomical signals. Lowering the detection 
thresholds to $\sim$2-3$\sigma$, in principle, should result in the detection of 
such signals, but this can be achieved only at the expense of a much larger number 
of false positives (mostly from signal statistics and perhaps some from RFI 
signals). As emphasised earlier, an underlying assumption is that the vast 
majority of these may be uncorrelated and therefore will be excised by coincidence 
filtering. In order to illustrate this, we conducted some specific analysis on 
suitably selected survey scans, the results from which are summarised below. 

\subsubsection{Case study: A field encompassing a known pulsar} \label{s:casetwo}

The detection of genuine astrophysical signals is illustrated through an example 
survey field that contains a known pulsar (PSR J1752$-$2806) at an offset of 
1.2$^{\circ}$ from the 
phase center (i.\,e.~$\approx 1.7 \times $ the half power beam width at 325 MHz). 
The strength of the signal is such that, at the incoherent array output, the 
brightest pulses from the pulsar mimic intermittent transient signals, thus 
providing a very good test case. We processed these data at the pulsar's DM 
(50.372 \dmu) and over the full recording bandwidth ($\Delta \nu$ = 16.66 MHz) 
as well as over a much reduced ($\Delta \nu / 8 \approx 2 $ MHz) bandwidth; the 
latter was done in order to emulate even weaker pulses. The resultant plots of 
pre- and post- coincidence filtering event rates are shown in Fig.~\ref{fig:psrdm}.

A quick inspection of these figures helps draw some useful conclusions. For example, 
in the full bandwidth case (left panel), all post-coincidence detection curves tend 
to merge near and above $\sim$1 $\sigma$, thus approaching the expected pulse rate 
$\approx$1.8 ${\rm s}^{-1}$. This may be interpreted as all genuine pulses that are 
bright enough (i.e. above the set detection thresholds) are detectable, thus providing 
crucial 
integrity checks of our processing pipeline. Secondly, for the reduced bandwidth 
case, where we emulate weaker pulses (i.\,e.~S/N $\approx$ 3 times lower), while the 
event rates for 2 or 3 sub-arrays at lower thresholds ($\la 1~\sigma$) are still 
dominated by false positives, a substantial improvement is seen on going to a 
larger number of sub-arrays. Overall this makes quite a compelling case to go for 
at least 4 sub-arrays. Furthermore, the improvement is only marginal on going from 
4 to 5 sub-arrays, which suggests that 4 sub-arrays may be an optimal choice for 
transient detection with the GMRT. 

\subsubsection{Detection of weaker signals} \label{s:foursubs}

Fig.~\ref{fig:foursubs} provides a useful illustration for the case of four 
sub-arrays.  We have taken a short stretch of data from the above field 
that contains a known pulsar and presented a time domain analysis. 
Of the 11 real pulses (transient signals) present in this short data block, only 6 are 
detectable with the sensitivity of the 30-antenna incoherent array and a 6$\sigma$ 
detection threshold (top panel). In order to ensure the detection of all pulses, it turns 
out that the detection threshold needs to be lowered to 3.5$\sigma$. As seen from 
the figure, this also results in many more false positives along side. A 3.5$\sigma$ 
threshold scales down to $\sim$1.8$\sigma$ when the array is sub-divided into 4 
distinct groups. Processing down to such low thresholds will obviously result in 
numerous false positives, as can be expected from signal statistics. However, as 
illustrated through this figure, virtually all of them are excised by the coincidence
filtering, resulting in a very small number of false positives in the end. In fact, a 
quick glance of the figure (lower most panel) reveals that all but the faintest pulse is 
detectable, along side a relatively smaller number of false positives compared to 
that of the full 30-antenna array.  This clearly illustrates that our basic theoretical
ideas proposed in \S~\ref{s:tech} do work in practice in real data. 

\subsubsection{Optimal number of sub-arrays} \label{s:allsubs}

Fig.~\ref{fig:allsubs} shows the net improvement achievable for different 
combinations of sub-arrays, i.\,e.~from 2 to 5, compared to a single incoherent 
sum of all 30 antennas as reference (top panel), for the data set used in the 
analysis above.  While there is a progressive 
improvement from 2 to 4 sub-arrays, the case for 5 sub-arrays is seen to be far 
less appealing compared to 4 sub-arrays. This inability of 5 sub-arrays to win 
over 4 sub-arrays may perhaps be due to possible departures in the detection 
sensitivity from the theoretically expected $\sqrt{\nant}$ for incoherent sum, 
or because the algorithm becomes less effective due to a larger number of false 
positives at such very low (1.5$\sigma$) thresholds. In short, 4 sub-arrays 
seems to be an optimal strategy for transient detection with the GMRT.

In order to quantify the level of improvements as well as to obtain more meaningful 
statistics, we processed the full duration of the scan (300 seconds) and conducted 
similar analysis, the summary of which is shown in Table 1. The improvements are 
tabulated both in terms of detections of real pulses as well as the number of 
false positives. The data were split into two halves for this analysis, as often 
the number of detections will critically depend on the modulation of pulse amplitudes 
(due to intrinsic and/or scintillation effects). The first four columns of the table 
are self-explanatory; column 5 is the fraction of the number of real events found (normalized to 
the total number of real pulses that are present in the data), and the column 6 the 
ratio of the number of false positives compared to that of the full 30-antenna 
incoherent array. Overall, the results are consistent between the two data sets, 
particularly the improvement factor in terms of the number of false positives. It is 
also evident that of all the combinations, 4 sub-arrays yields the best improvement, 
which supports our finding from the example illustrated through Fig.~\ref{fig:allsubs}.


\section{Event analysis pipeline} \label{s:event} 

Promising candidate events that emerge from our processing pipeline are subjected 
to detailed scrutinies, a basic scheme for which is shown in Fig.~\ref{fig:eventpipe}. 
Among the salient features are the integration of a calibration and imaging pipeline 
for potential on-sky localization of the event and the ability to phase up the full 
array toward the target position to enable a high-quality signal detection and 
confirmation. This necessitates keeping track of the calibrator observations and 
processing them routinely to solve for the complex gains of the array elements. For 
localization via imaging, the raw data segments of an event are first correlated 
to generate visibility data.   Performing phase-coherent dedispersion prior to 
correlation can greatly increase the chances of localization, especially for 
short-duration signals at moderate or high DMs (e.g. giant pulses). In the event 
that a clear detection in imaging and accurate localization ($\sim$5-10'') is possible, 
a sensitive phased-array beam can be formed toward the target position to enable 
a high-quality signal detection and characterisation\footnote{As the localisation 
radius scales as $(S/N)^{-1}$ for unresolved point sources,  accuracies at the level
of an arc second are achievable even in the case of marginal ($\sim$5-10$\sigma$) 
detections at 
325 and 610 MHz.}. Depending on the characteristics of the signal (e.g. time 
duration, temporal structure and DM), the phased-array data can then be processed 
for phase-coherent or incoherent dispersion removal followed by detection and 
subsequent analysis and further checks. As well as enabling crucial integrity 
checks of the detected events, such a powerful methodology offers the advantages 
of obtaining additional information -- such as high time resolution studies, 
accurate DM estimation and localization of the target -- for any genuine signals 
that may need further detailed follow-ups. Some of these possibilities are further 
elaborated and illustrated through suitable examples in the subsequent sections.

 
\begin{figure*}
\epsscale{2.0}
\plotone{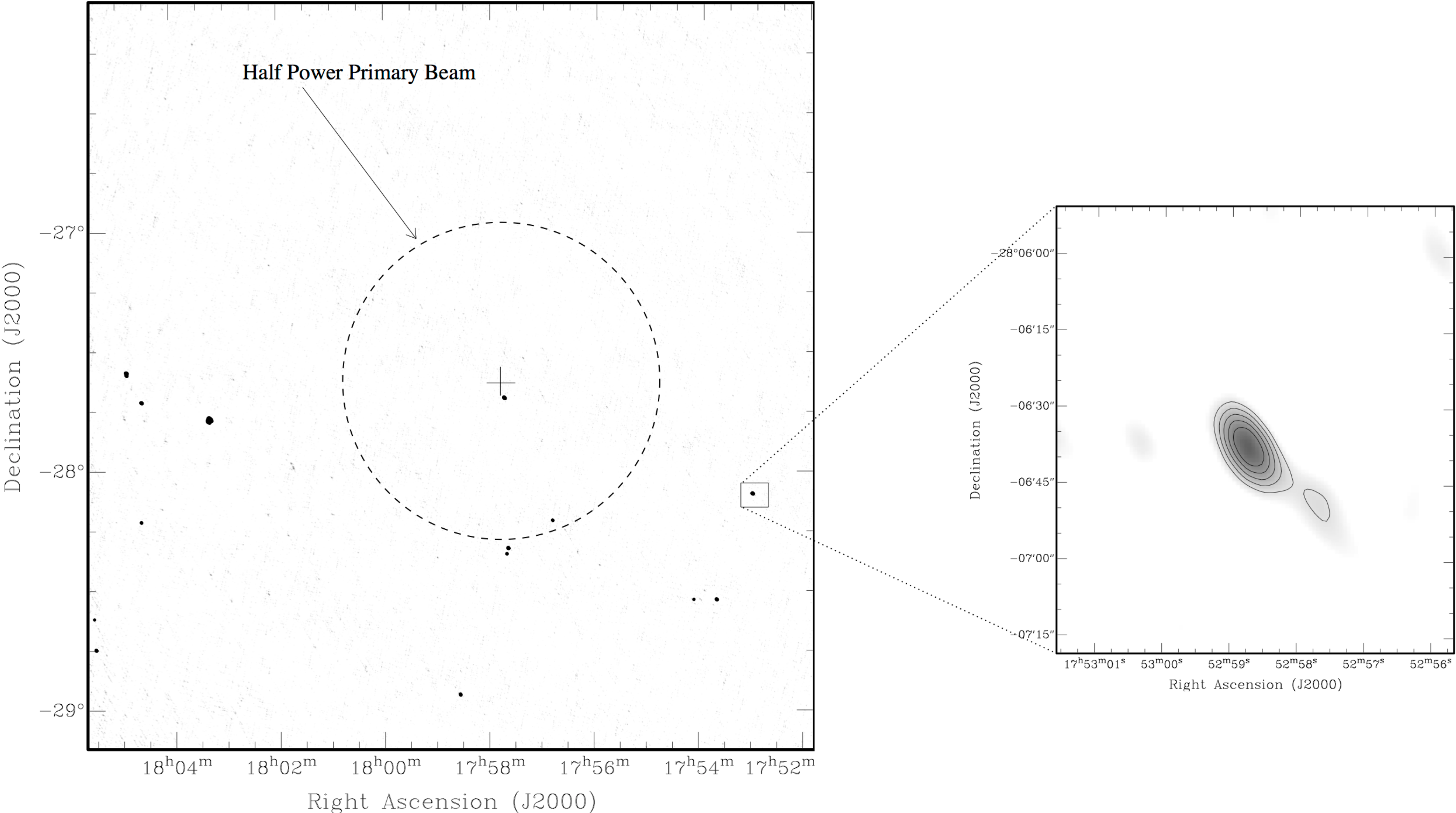}
\caption{ Image made from the correlated visibility data for the pointing 
GTC\_002.52$-$1.64.
The total time interval over which the visibilities are calculated is 0.8 
seconds. The pulsar J1752$-$2806 was detected in this pointing by the transient
search pipeline. The image shows a number of sources in the field, and the source 
corresponding to J1752$-$2806  is marked with a box. The image with the large 
field-of-view has a resolution of 36" x 20". A zoomed-in image of the 
pulsar emission with a resolution of 16.5" x 8.5" is also shown. The 
contour levels in the zoomed in image start at 41 mJy and are in steps
of 9 mJy. The position and flux measured from this image is 
RA = 17 52 58.746  $\pm$ 0.024, DEC = --28 06 36.09 $\pm$ 0.41 
and $80 \pm 9$ mJy.}
\label{fig:gtc252image}
\end{figure*}


\subsection{Imaging pipeline} \label{s:imaging}

As described above, one of the major advantages of transient detection via 
interferometric arrays is the possibility of localisation of the source. This
is most straightforwardly done by making an image of the transient. As
described elsewhere in this paper, once a particular data stretch has been
identified as containing a possible transient, the voltage data from each
antenna for that corresponding time interval is saved. This data is then
correlated (using essentially the same correlation routines as used in
the real time system) to create a set of visibilities. The process of making
an image from these visibilities is well understood (see e.g. \citet{thompsonetal2001}), 
and there exist several software packages aimed
at doing this problem (e.g. AIPS, Miriad, CASA). The principal steps are
(i) identifying and flagging out erroneous visibilities, e.g. those affected
by radio frequency interference, which can be significant at most of 
the frequencies at which the GMRT operates, (ii) correcting for the 
complex gain (including the atmospheric/ionospheric gain) and (iii) imaging
and deconvolution. The first two of these steps have been incorporated into
a pipeline \flagcal \citep{flagcal}, while the imaging and deconvolution 
is currently done using one of the standard packages (AIPS in this instance)

\subsubsection{Identification and flagging of corrupted visibilities} \label{s:flag}

The most common type of strong RFI at the GMRT site has a small occupancy in
the time-frequency space, i.\,e.~is either limited in time, or in 
frequency, or in both. \flagcal uses this fact to identify corrupted visibilities.
Essentially robust statistics (across time, frequency and baselines) of the 
visibilities are derived, and then outliers with respect to these statistics
are identified and flagged out. Slow variations in the visibilities are
accounted for by allowing for a (user definable) smoothing in the time
frequency plane before computation of the statistics and identification
of the outliers. In calibrator scans, one would expect that, in the absence
of any corruption, the phase of the visibility would be nearly constant
on the typical timescale of a calibration observation (i.\,e.~of the order of 
a few minutes).
This is also used to identify corrupted data. RFI often affects contiguous
sets of channels and or time ranges, and hence two passes are made through
the data, one of which identifies corrupted visibilities on the basis of
the robust statistics, and the other that marginalises over the flagged 
data to identify frequency channels, baselines and/or antennas for which the data 
has been corrupted. The output of this stage of the pipeline is a set of
visibilities in which all data identified as being corrupted has been flagged
out. Since the determination of robust statistics is computationally intensive,
\flagcal implements this using OpenMPI, resulting in significant speed ups
in multi-core machines.


\begin{figure*}[t]
\epsscale{2.0}
\plotone{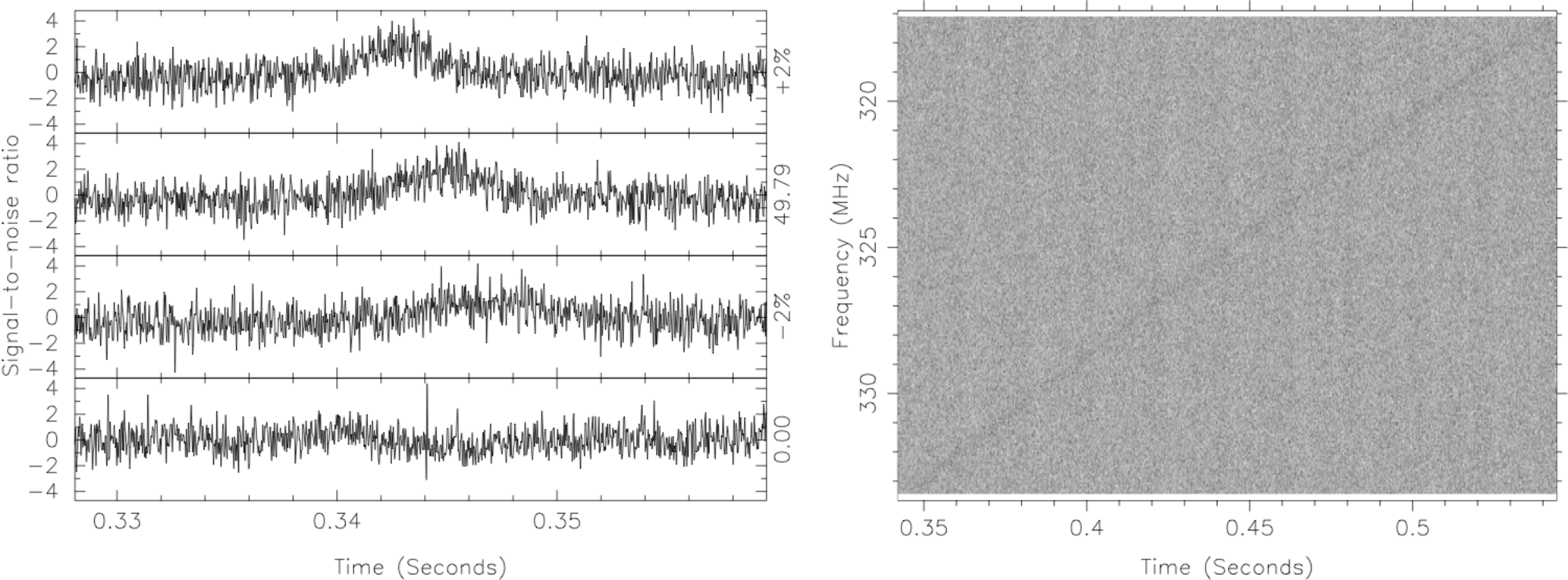}
\plotone{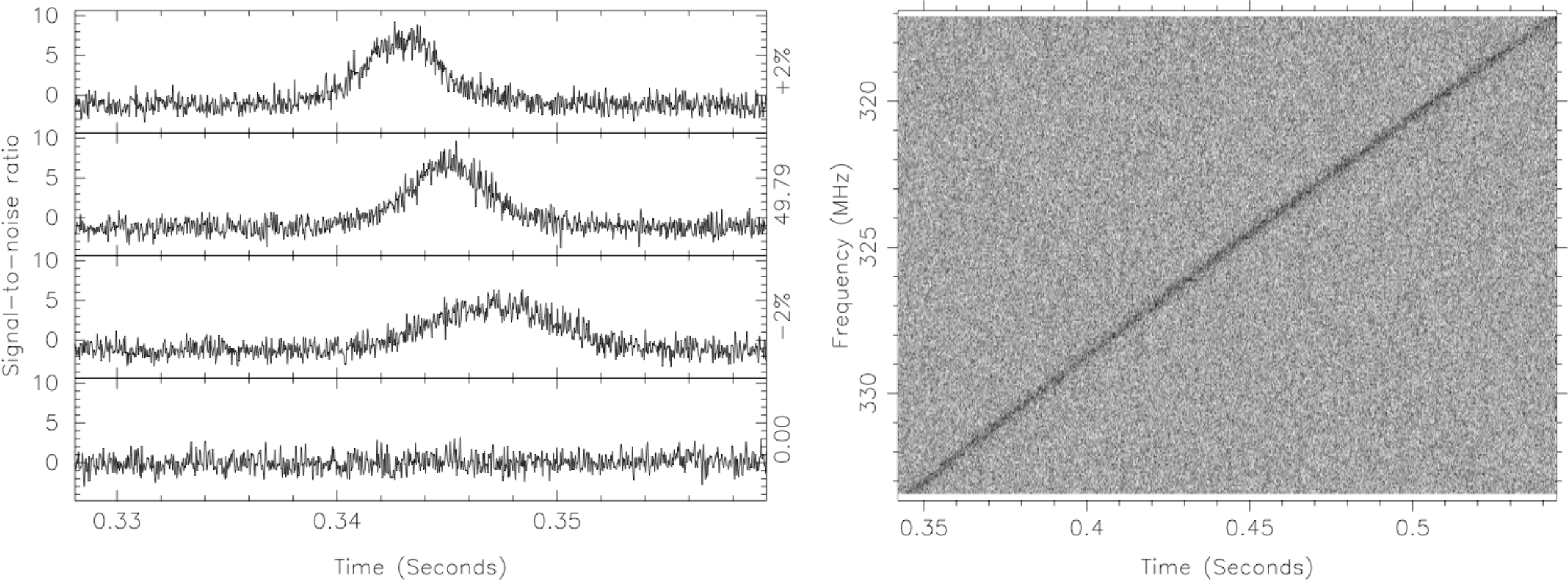}
\caption{Example data illustrating the detection of a candidate signal and its 
subsequent confirmation and verification. The transient pulse (from a known pulsar 
that was within the FoV) was first detected at a relatively low 
significance ($\approx$5$\sigma$ at $\approx$0.25 ms resolution) and was treated 
as a candidate as it passed the set coincidence criteria. A sensitive phased array was 
then formed by phasing up all antennas with good calibration solutions toward the 
source position as determined by the on-sky imaging pipeline (see Fig.~\ref{fig:gtc252image}). 
The improved signal detection from this phased-array data is shown in the bottom panels. 
The peak signal-to-noise ratio improved by a factor of six, which is approximately 
70\%  of the expected level of improvement given only 21 of the 30 antennas were 
phased up for final detection and verification. } 
\label{fig:iapa}
\end{figure*}

\subsubsection{Calibration} \label{s:calib}

At any instant, an N-element interferometric array (i.\,e.~one in which there
are N unknown complex instrumental gains) measures $N (N-1) / 2$ complex
visibilities. This makes the problem of determining the antenna gains from
observations of a calibrator source with known visibilities (e.g. a point
source at the phase centre) over determined. Iterative schemes for determining
the least squares solution for this problem have been described by e.g. \citet{bhatnagar2001}.
\flagcal implements this iterative scheme to determine the complex antenna 
gains. In general there are several kinds of calibrations that can be performed, {\it viz.} 
``flux calibration'' for determining the absolute flux level; ``bandpass calibration'' for 
the spectral response; and ``phase calibration'' for the combination 
of the atmospheric and instrumental gains.  \flagcal implements all of these
calibrations and also interpolates the final corrections on to the target
visibilities. It also allows interpolation of the flags from the 
calibration scans onto the target visibilities. This is useful in (the
commonly encountered) situation where there is persistent RFI affecting some
given spectral channels, antennas or baseline combinations.


\subsubsection{Imaging and Deconvolution} \label{s:deconv}

The flagged and calibrated visibilities computed by \flagcal are written out
as a FITS file. This allows easy processing using standard imaging packages.
This stage of the processing can easily be automated. The GMRT array 
configuration has been designed to give a fairly good snapshot UV coverage 
at most declinations (see \citet{swarupetal1991})
allowing for a good localisation of the source. As described above, this
localisation is also important in determining that the transient emission
that was detected indeed arises from the sky. In situations where there is 
sufficient signal-to-noise ratio, further confirmation of this comes from
confirming that self calibration improves the signal-to-noise ratio of
the image.

\subsubsection{Case Study PSR J1752$-$2806} \label{s:case}

The pulsar J1752$-$2806 was targeted as part of the test observations on 2010 
February 20. Calibration was done using the data from the source 1830$-$360 from 
the VLA catalog. The data was run through the \flagcal pipeline, and then imaged 
using the AIPS task IMAGR. The image was produced by excluding the edge channels, 
as well as some central channels which were badly affected by RFI and for which 
most of the data was flagged out. The final bandwidth used to make the image was 
$\sim 14.7$~MHz (i.\,e.~$\approx$90\% of the recording bandwidth), and the noise 
level in the image is $\sim 12$~mJy. The pulsar is clearly detected at a flux level 
of $\sim 220$ mJy. A strong confirmation that the emission arises from the sky (and 
is not some chance RFI) comes from the fact that  self calibration significantly 
increased the peak flux of the source; specifically, the flux after one round of 
phase only self calibration is $\sim 280$ mJy. A similar confirmation comes from 
the fact that cleaning (which was done using the AIPS task IMAGR) substantially 
reduces the sidelobe levels. Fig.~\ref{fig:1748image} shows the images produced 
before and after cleaning.  


\subsection{Application for event localization} \label{s:snap}

We present another case study where a transient pulse was blindly detected in our 
search pipeline. The scenario is the same as that outlined in \S~\ref{s:casetwo}, 
i.e. a survey field that contained a known pulsar but at a large offset of 
$\approx$71$^{\prime}$ from the phase centre. This large offset (1.7 $\times$ 
the nominal half power beam width) means a source location near the edge of the 
beam and as a result the pulsar will effectively be detected as an intermittently 
emitting  transient source. The pulse was detected as a 5$\sigma$ event in the 
search pipeline and the signal characteristics (i.e. arrival time and DM) were 
then used to determine and extract the corresponding raw data segments from all 
30 antennas. These data were then correlated to produce the visibilities which 
was subsequently imaged using the procedures described in \S~\ref{s:imaging}. 
Observations of 1830$-$360 that was recorded 36 minutes prior to the detection 
time of the transient pulse were used for calibrating the visibilities.  

As a demonstration of our event localisation strategy, a snap-shot image was made of 
a $ 3^{\circ}\times 3^{\circ} $ region (nominal full beam width $\sim$1.4$^{\circ}$) 
of the sky centred at the phase centre of the survey pointing (RA = 17h 57m 51.48s, 
DEC = $-$27d 36' 00.0"). The pulsar was clearly detected in the image along 
with several other point sources in the field (see Fig.~\ref{fig:gtc252image}). 
The estimated pulsar position of RA = 17 52 58.746 $\pm$ 0.024, 
DEC = $-$28 06 36.09 $\pm$ 0.41 is within 2-3$\sigma$ of the catalog 
position\footnote{The actual positional uncertainties will be of the order 
of one third of the beam size, i.e. approximately 0.3 s in RA and 3" in DEC.} 
and the measured flux $\sim$ $80\pm9$ roughly agrees with the expected flux 
after scaling for the primary beam. 

It is worth noting that even at this relatively bright flux level,
there are a number of sources within the FoV.  A cross check
of the source positions in the image with the NVSS catalogue shows 
a good correspondence. The large number of ``confusing" sources may
partly be because the target field is close to the Galactic
centre ($l = 1.5^{\circ},~b= -1^{\circ}$). At fainter flux
levels however, one would expect that there would be a number of 
background sources that would be present in the FoV. To
distinguish between these and the transient source, we may apply either
of the following two strategies: (i) make a fresh image centred on the 
time range during which the transient was the brightest -- presumably
the only source in this image whose flux will vary will be the transient;
or, (ii) redo the transient search with a phased-array beam centred on 
each of the candidate sources -- the signal-to-noise ratio would be the 
largest when the antennas are phased up toward the right position.


\subsection{Phased array for improved signal detection and confirmation} \label{s:phasing}

In addition to producing incoherent array beams, the GSB beamformer can also generate 
coherent (phased) array beams. This involves performing suitable addition of 
pre-detected voltage samples from individual antennas. Coherent beam formation 
however requires calibrating out  the antenna based phase offsets before the voltage 
samples can be added.  These antenna based phases are solved using the recorded 
cross-correlations on a calibrator source near the target position, typically 
observed alongside the observations . As outlined in \citet{royetal2010}, these 
phases are applied after the FFT stage, as an additional term in the fringe corrections. 

As discussed in \S~\ref{s:tech}, the incoherent array beam has the same field-of-view 
as the primary beam of a single antenna, but with an enhanced sensitivity of 
$\sqrt{N_{a}}$ times that of a single antenna, for an array of $N_{a}$ antennas.  
However, the coherent array beam is much narrower than that of a single antenna -- 
similar to the synthesized beam obtained from the array of $N_{a}$ antennas, and 
therefore results in a  sensitivity improvement of $N_{a}$ times than that of a 
single antenna. Hence by forming the coherent array beam towards the target source 
after phasing up the array, we expect $\sqrt{N_{a}}$ sensitivity improvement 
compared to the incoherent array. 

As a demonstration of the follow-up strategy outlined above, we formed a 
phased-array beam at the position of the transient pulse in Fig.~\ref{fig:gtc252image}. 
The pulse was detected at a significance of 5$\sigma$ at a time resolution of 
$\approx$0.25 ms.\footnote{Even though a higher significance is possible via 
matched filtering, we limit the time resolution to 0.25 ms for the purpose of 
this analysis.} The initial detection (at this resolution) and the final detection 
from processing the phased-array data are shown in Fig.~\ref{fig:iapa} 
(top and bottom panels respectively). Data on the same calibrator source 1830--360 
(i.e. recorded $\sim$30 min prior to the detection of the pulse) were used to solve 
for the antenna based phases required for phasing up the array. As seen from the 
figure there is 6 times improvement in the signal-to-noise ratio compared to the 
initial detection from the search pipeline. This is almost 70\% of the theoretically 
expected improvement. \footnote{Only 19 of the 30 antennas were phased up for the 
final detection. Antennas with poor phase solutions were flagged from the analysis 
to maximise the signal detection.} A similar analysis was conducted on multiple 
other pulse detections and it suggests that up to $\sim$80\% of the theoretically 
expected improvement may be achievable in practice. Even so, significant S/N 
improvements (as much as a factor 10) are still achievable in the final detections. 

The discrepancy may be attributed to plausible calibration inaccuracies or some 
possible dephasing of arm antennas (due to ionospheric effects) given the 
12.42$^{\circ}$ separation between the pulsar and calibrator positions. In-beam 
calibration may help alleviate this in principle, however the GMRT's FoV 
may often limit its prospects; e.g. while there are multiple point sources in 
Fig.~\ref{fig:gtc252image}, the brightest source has a flux of only $\sim$100 mJy, 
not good enough to derive reliable calibration solutions. However, this will no 
longer be a limitation for future wide-FoV instruments such as MWA, LOFAR and 
ASKAP that will contain multiple potential calibrator sources in any given field. 


\section{Future Work} 

While the analysis presented in this paper is largely based on our pilot survey 
data at 325 MHz, it is important to ascertain the efficacies and limitations of 
conducting transient searches at different frequencies of the GMRT. For instance, 
the RFI environment varies significantly between different frequencies and it will 
be very useful to investigate the effectiveness of snap-shot imaging for developing 
immunity against a variety of RFI-generated events. This will be the subject of a 
future publication.  While the lower frequencies provide the basic advantage of 
comparatively larger fields-of-view (e.g. $\sim$6 ${\rm deg^2}$ at 150 MHz 
compared to $\sim$1.5 ${\rm deg^2}$ at 325 MHz), they also imply increased 
challenges in terms of having to deal with more 
severe RFI environments. The higher frequencies (e.g. 1400 MHz),  on the other hand, 
while helping to extend the parameter space (i.e. searching out to larger DMs), may 
necessitate trading-off the achievable field of view ($\sim$0.1 ${\rm deg^2}$). We 
will conduct similar pilot surveys for all other frequencies GMRT to further optimise 
our processing pipelines and search algorithms. 

Our eventual goal is a transient detection system for the GMRT that functions in a 
commensal mode with other observing programs. Having demonstrated the efficacies 
of multiple (incoherent) sub-arrays for initial detection and interferometric 
capabilities for on-sky localization, the next logical step is a real-time 
implementation of such a pipeline. The GPU-optimized dedispersion software 
developed at Swinburne \citep{barsdelletal2012} has been benchmarked for the 
GMRT frequencies and the current recording bandwidth of 32 MHz, and can 
handle up to $\sim$550 trial DMs (at both 325 and 610 MHz). The matched-filtering 
based detection is relatively inexpensive computation-wise and can easily be integrated, 
however recovering the loss in sensitivity (due to sub-arraying) through the use of very 
low detection thresholds (i.e. $\sim$2-3$\sigma$ compared to $\sim$5-6$\sigma$ 
typically used in transient searches) may require some optimisation of the downstream 
algorithms for event scrutiny. 

We have outlined and demonstrated a specific approach for transient detection 
with interferometric arrays. While we advocate the use of multiple incoherent 
sub-arrays and coincidence checks as a promising strategy, there may be other 
possibilities that are worth exploring within the general context of next-generation 
instruments such as ASKAP, MeerKAT and the SKA, including, for example,
the use of multiple sub-arrays to achieve larger fields-of-view (i.e. by pointing 
the sub-arrays in different regions of the sky). The improved sensitivity 
achievable through the use of wider-bandwidth recorders, and the constraints that 
may arise in terms of data rates and processing needs, are also among important 
aspects that need investigation. As the GMRT gets upgraded in the coming years 
through commissioning of its broad-band receivers and backends, new avenues will 
be opened up for undertaking such more promising, albeit more complex, science 
demonstrator projects relevant in the SKA-era.


\section{Summary and Conclusions}

While large single-dish instruments currently dominate time-domain science applications
such as pulsars and fast transients, 
the future lies in the effective use of large-element interferometric arrays. The GMRT, 
with its modest number of elements and long baselines, makes a powerful platform for 
developing the necessary techniques and methodologies. In particular, its sub-array and 
interferometric capabilities can be well exploited for efficient detection of fast transients
as well as for their accurate on-sky localisation. 

Among the various considerations in the use of arrays for transient exploration 
is the trade-off between the field of view and absolute detection sensitivity. 
We have postulated the basic idea of generating a relatively small number
of incoherently summed sub-arrays from the full array and then combining the results
of detections of candidate transient events from each of these sub-arrays so as
to optimise the rejection of false positives due to receiver noise and from RFI, 
using suitably devised coincidence filtering techniques. As we demonstrate
through multiple examples and analysis, this enables reaching the sensitivity of 
the full phased array, while preserving the full FoV of the single antenna element. 
This approach is promising as it offers a dramatic improvement in terms of the prospects 
of detecting weaker signals. For example, a $\sim$2$\sigma$ detection from initial 
processing will eventualy  be a $\sim$10$\sigma$ signal after phasing-up the full array, 
and hence will be both unambiguously verifiable in time series as well as localizable 
(on sky) at arc second accuracies. 

The GMRT software backend allows raw voltage data from individual array elements 
to be recorded and made available for software-based pipelines. We have exploited this 
optional feature to develop and implement a transient detection pipeline for the GMRT. 
This includes a beamformer that operates on 2 $\times$ 30 raw voltage data streams to 
produce multiple incoherently summed sub-arrays, the data from which are then 
dedispersed and searched for transient events. The resultant events are scrutinised by 
the coincidence algorithms that take into account likely differences in the detection 
sensitivity between different sub-arrays that may result from either local RFI, or from 
one or more array elements performing at less than their nominal sensitivities. We have 
also explored the effectiveness of the algorithms as a function of the detection threshold 
as well as sub-arraying, and our analyses suggest that four sub-arrays make an optimal 
choice for  transient detection with the GMRT. 

Important future work includes undertaking pilot surveys at different frequencies of the GMRT 
in order to further optimise the strategies and algorithms for transient detection with 
arrays, and the development of a real-time version of the transient pipeline that can
work in commensal mode. 

The work described here, while demonstrating the applications of interferometric 
arrays for fast transient exploration -- an important preparatory step for planned 
science with the SKA pathfinder instruments such as ASKAP and MeerKAT -- 
also forms the first step to add new capabilities to the GMRT in the exciting arena 
of charting the transient sky at radio wavelengths.

\bigskip

\noindent
{\it Acknowledgements:} 
This work is supported by the Australian Government under the Australia-India Strategic Research Fund grant ST020071
and by the Indian Government under the Department of Science and Technology grant DST/INT/AUS/Proj-14/2008. 
The Centre for All-sky Astrophysics is an Australian Research Council Centre for Excellence, funded by CE11E0090. 
N.D.R.B. is supported by a Curtin Research Fellowship and thanks Steven Tingay for support and encouragement 
extended to this project. We thank an anonymous referee for many useful comments that helped to improve the clarity 
of the paper. Data processing was carried out at Swinburne University's Supercomputing Facility. We thank the staff 
of the GMRT for help with the observations.  The GMRT is operated by the National Centre for Radio Astrophysics 
(NCRA) of the Tata Institute of Fundamental Research (TIFR), India. 


\begin{figure}
\epsscale{1.0} 
\plotone{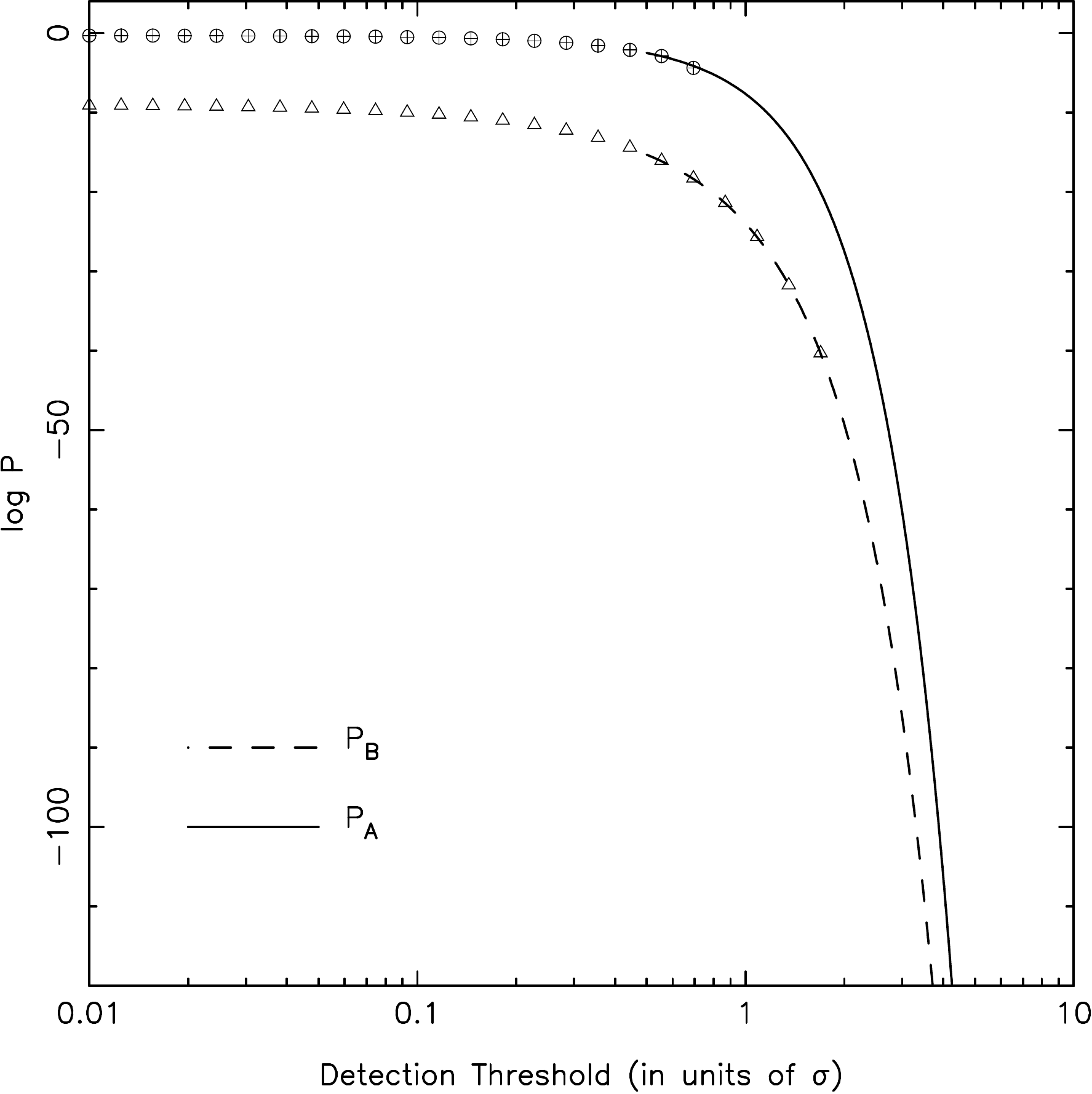} 
\caption{The change in the computed PFA with threshold $r$ (in units of $\sigma$) for 
the case of one IA beam with $N$ antennas in the sub-array (solid line) and for the case
of  $N$ number of IA beams with one antenna in each sub-array (dashed line).  The 
symbols mark the results from numerical simulations using Gaussian random noise.}
\label{fig:prob}
\end{figure}

\section{Appendix: Probability of False Alarms for Sub-Array configurations} \label{appendix} 

For the case where the random noise output signal of a radio telescope antenna follows 
a Gaussian distribution with zero mean and variance of $\sigma^2$, the probability 
of the signal level crossing a threshold $T$ is given by:

\begin{equation}
P(>T) = \int_{T}^{\infty} \frac{1}{\sqrt{2\pi} \sigma} \exp[-\frac{x^2}{2 \sigma^2}] dx 
\end{equation}

or 

\begin{equation}
P(>T) = \int _{T/\sqrt{2}\sigma}^{\infty} \frac{1}{\sqrt{\pi}} \exp[-y^2] dy = \frac{1}{2}
{\rm Erfc}\left(\frac{T}{\sigma \sqrt{2}} \right) 
\end{equation}

where ${\rm Erfc}$
is the complimentary Error Function.  We call $P(>T)$ the probability 
of false alarms (PFA), as these excursions would lead to false triggering of a transient
detection pipeline where the detection threshold is set to $T ~ \sigma$.

For an array of $N$ such antennas, the signal can be combined in different ways.  For 
a coherent phased array, the voltage signals from individual antennas are added in phase
and then squared to get the total intensity, and then further integrated in time and 
frequency as required to get the final output.  For an incoherent array, the intensity 
signals from individual antennas are added to get the total intensity signal and then
integrated to the desired time and frequency resolution.  The effective $\sigma$ 
decreases by a factor of $N$ or $\sqrt{N}$, for the coherent and incoherent array
outputs, respectively \citep{guptaetal2000}.
Thus, for an incoherent array (IA) of $N$ antennas, the effective sigma is :

\begin{equation}
\sigma_N = \frac{\sigma}{\sqrt{N}}
\end{equation}
where $\sigma^2$ is the variance for a single antenna.

In what follows we consider the following three cases of incoherent array :

\noindent
{\bf\ (A) Single IA beam from a single sub-array of $N$ antennas:}

For this case the probability of false alarms can be expressed as (from eqn A1 above):

\begin{equation}
P_{A}(>T) = \frac{1}{2}\mbox{Erfc}\left( r \sqrt {\frac{N}{2}}\right)
\label{casea}
\end{equation}
where $r ~=~ T/\sigma$ is the detection threshold in units of $\sigma$.

{\bf\ (B) $N$ IA beams from $N$ sub-arrays, each with one antenna:}

In this case the false alarms due to noise statistics are independent events 
in each sub-array output, and if we use a coincidence filtering scheme where
a false alarm is declared only if it is present simultaneously in all sub-array
outputs, then the PFAs from all the IA beams get multiplied to give 

\begin{equation}
P_{B}( > T)  = \left [\frac{1}{2}\mbox{Erfc}\left( \frac{T}{\sigma \sqrt{2}} \right) \right]^N
\end{equation}


\noindent
{\bf\ (C) $p$ IA beams from $p$ sub-arrays, each with $n=N/p$ antennas:}

The most general case is to have $p=N/n$ sub-arrays with each having $n$ antennas, 
for which the effective PFA is given by 

\begin{equation}
P_{C}(>T)  = \left [\frac{1}{2}\mbox{Erfc}\left(\frac{T}{\sigma} \sqrt{\frac{n}{2}}\right) \right]^p
\end{equation}

Note that for $p=1,n=N, P_C=P_A$ and $p=N,n=1, P_C=P_B$.  Also, it is easy to show
that, for a given threshold $r={T}/{\sigma}$, $P_B < P_C < P_A$.   This is illustrated
in Fig.~\ref{fig:prob}, which plots the PFA as a function of $r$ for the two cases $P_B$ and $P_A$, 
for the GMRT value of $N~=~30$.  Results from simulation runs using random Gaussian noise 
(denoted by the symbols) are also overploted on the theoretically calculated curves. 
As can be seen, for small threshold values, log$P_B$ is less than log$P_A$ by a factor 
of 10 and this ratio increases with $r$ for values of $r ~>~ 1$, reaching a value of
around 26 for a threshold of 3.0.  The curve for $P_C$ would lie in between the curves
for $P_A$ and $P_B$.  

From this, it is evident that there exists a 
possibility for trading off between PFA and $r$, for different choices of sub-arrays.
For example, in order to have the same PFA for different combinations of sub-arrays,
it is possible to work at lower thresholds for cases where the array is split into
sub-arrays.  This can offset the basic reduction in sensitivity that each sub-array
suffers from (compared to the case of a single sub-array of $N$ antennas), while 
offering improved immunity against local interference signals that don't pass the
coincidence filtering test.

As an alternate illustration of the ideas, fig A2 shows the ratio of $P_C$ to $P_B$ 
(on a log scale) as a function of number of antennas in the sub-array, $n$,
for different choices of total number of antennas, $N$, and for a fixed choice 
of $r~=~3.0$.  Moving along any of these curves from $n~=~N$ to $n~=~1$ illustrates
how the PFA reduces as more numbers of sub-arrays are used. 

Thus, it is possible to reduce the probability of false alarms by using multiple
sub-arrays, and this can be used to trade-off sensitivity (via different threshold 
values) vs false alarm rate to optimise the performance from the entire array.



\begin{deluxetable}{lccccc}
\tablewidth{20pc}
\tablecaption{Coincidence filtering: event list summary \label{tab:sum}}
\tablehead{
\colhead{$N_{\rm sub}$} & \colhead{$r$} & \colhead{$N_{\rm real}$} &   \colhead{$N_{\rm false}$} &   \colhead{$f _{\rm real}$} &  \colhead{$f _{\rm false}$} \\
\colhead{ (1) }  & \colhead { (2) } & \colhead { (3) } & \colhead { (4) } & \colhead { (5) } & \colhead { (6) } 
}
\startdata
   &         &            &               &              &           \\
\multicolumn{5}{l}{Data segment 1 (0 - 150 seconds) } \\
   &         &            &               &              &           \\
1 & 3.5   & 62      & 1193     & 0.25        &    1          \\
2 & 2.5   & 113    & 615       & 0.45        &     0.52    \\
3 & 2.0   & 121    & 653       & 0.49        &    0.55     \\
4 & 1.7  & 112    & 367       & 0.45        &     0.31    \\
5 & 1.5  & 108    & 537       & 0.43        &     0.45    \\
   &         &            &               &              &           \\
\multicolumn{5}{l}{Data segment 2 (150 - 300 seconds)  } \\
   &         &            &               &              &           \\
1 & 3.5   &  92    &  1243 &      0.39    &         1   \\
2 & 2.5   & 146   &  601   &    0.62      &       0.48  \\
3 & 2.0   & 119   &  761   &     0.51     &        0.61  \\
4 & 1.7   & 133   &  393   &     0.57     &        0.32   \\
5 & 1.5   & 131   &  564   &     0.56     &        0.45  \\
\enddata
\tablecomments{Results from a coincidence filtering analysis of the survey field GTC\_002.52--1.64
where a known pulsar (PSR J1752$-$2806) was present at an offset of $\approx$1.2$^{\circ}$ from 
the beam phase centre (i.e. outside the half power primary beam; see Fig.~\ref{fig:gtc252image}).
Number of sub-arrays ($N_{\rm sub}$) varied from one to five, and the detection thresholds are
scaled down assuming the theoretically expected $\sqrt{n}$,  where $n$ is the number of antennas
per sub-array; the number of real events and false positives ($N_{\rm real}$ and  $N_{\rm false}$) 
are tabulated along with the corresponding fractions ($f _{\rm real}$ and $f _{\rm false}$).}
\end{deluxetable}

\bibliographystyle{apj}

\end{document}